\documentclass[fleqn,usenatbib]{mnras}
\usepackage{graphicx}
\usepackage{natbib}
\usepackage{epstopdf}
\usepackage{amssymb}
\usepackage{amsmath}
\usepackage{mathptmx}
\usepackage{multirow}
\usepackage{array,booktabs}
\usepackage{hyperref}
\usepackage{float}
\usepackage{subfig}

\newcommand{\msun}{{\rm~M_\odot}}

\newcommand{\llgrbs}{{{\it ll}GRBs }}

\newcommand{\appropto}{\mathrel{\vcenter{
  \offinterlineskip\halign{\hfil$##$\cr
    \propto\cr\noalign{\kern2pt}\sim\cr\noalign{\kern-2pt}}}}}

\title[Jet Driven Explosion]{Observational signatures of stellar explosions driven by relativistic jets}
\author[Eisenberg, Gottlieb \& Nakar]{
    Moshe Eisenberg$^1$
	Ore Gottlieb$^2$,
	Ehud Nakar$^1$ \thanks{udini@tauex.tau.ac.il}
	\\
	{$^1$School of Physics and Astronomy, Tel Aviv University, Tel Aviv 69978, Israel}\\
	{$^2$Center for Interdisciplinary Exploration \& Research in Astrophysics (CIERA), Physics \& Astronomy, Northwestern University, Evanston, IL 60201, USA}
}
\pubyear{2021}
\begin{document}
	\label{firstpage}
	\pagerange{\pageref{firstpage}--\pageref{lastpage}}
	\maketitle	

\begin{abstract}
The role of relativistic jets in unbinding the stellar envelope during a supernova (SN) associated with a gamma-ray burst (GRB) is unclear. To study that, we explore observational signatures of stellar explosions that are driven by jets. We focus on the final velocity distribution of the outflow in such explosions and compare its observational imprints to SN/GRB data. We find that jet driven explosions produce an outflow with a flat distribution of energy per logarithmic scale of proper velocity. The flat distribution seems to be universal as it is independent of the jet and the progenitor properties that we explored. The velocity range of the flat distribution for typical GRB parameters is $\gamma\beta \approx 0.03-3$, where $\gamma$ is the outflow Lorentz factor and $\beta$ is its dimensionless velocity. A flat distribution is seen also for collimated choked jets where the highest outflow velocity decreases with the depth at which the jet is choked. Comparison to observations of SN/GRBs rules out jets as the sole explosion source in these events. Instead, in SN/GRB the collapsing star must deposit its energy into two channels – a quasi-spherical (or wide angle) channel and a narrowly collimated one. The former carries most of the energy and is responsible for the SN sub-relativistic ejecta while the latter carries 0.01-0.1 of the total outflow energy and is the source of the GRB. Intriguingly, the same two channels, with a similar energy ratio, were seen in the binary neutron star merger GW170817, suggesting that similar engines are at work in both phenomena. 
\end{abstract}

\vskip 0.5cm


\section{Introduction}
Massive stars end their lives in a powerful supernova (SN) following the collapse of the stellar core. While the general picture of core-collapse SNe (CCSNe) is known, the mechanism(s) powering the shocks that drive SNe is not clear yet. Models suggested include some combination of neutrino heating, magnetohydrodynamic power, accretion power, thermonuclear burning, shock bounce and acoustic mechanisms (see e.g., \citealt{janka2012} and references therein). The most popular model, at least for the most common type II SNe, is neutrino-driven explosion, possibly assisted by hydrodynamic instabilities. Yet, it is almost certain that at least some SNe are driven by other mechanism(s) (see below).

In some massive stars, whose properties are not fully understood, the collapse of the core also generates a long gamma-ray burst (LGRB) together with the SN. The origin of the LGRB is an ultra-relativistic (Lorentz factor $\gtrsim 300$) narrow jet that carries a minute fraction of the progenitor mass ($\lesssim 10^{-6}\msun$). By contrast, the SN emission is produced by a massive  ($\sim 10\msun$) sub-relativistic ($\sim 10,000$ km/s) ejecta. While we do not know what the exact geometry of the SN ejecta is, it is clearly much closer to spherical than to narrowly collimated relativistic jets. The CCSNe that are associated with LGRBs are of the rare broad-line Ic (Ic-BL) type \citep[][ and references therein]{woosley2006}. Namely, the ejecta is H and He poor and its velocity is relatively high ($\gtrsim 10,000$ km/s; hence the broad lines). SNe that are associated with LGRBs are typically extremely energetic, carrying more than $10^{52}$ erg in kinetic energy. This energy is high not only when compared to other SNe, it is also higher by a factor of 10-100 (or even more) compared to the energy carried by the relativistic jet in those events (see e.g., \citealt{mazzali2014}, and references therein).
To conclude, the collapse of the core of an LGRB/SN progenitor generates two outflows, a more energetic massive slow and wide angle outflow that radiates the SN light and a less energetic low-mass relativistic narrow jet that is the source of the GRB. The question of how the collapse of a single core generates both outflows is still open. We know that a fast rotating compact object (black hole or neutron star) must be formed in order to produce the jet. It is also highly unlikely that the SN is driven by the popular neutrino-driven mechanism, since the SN energy is too high. But, we do not know what the energy source of the SN is and whether it is related to the GRB jet. Moreover, it is still unknown if both outflows are generated simultaneously, and if not then which of the two is the first and what is the time separation. Similarly, we do not know what the interplay between the two components is, if there is any \citep[e.g.][]{de-colle2021}.  

Apart from the interest in the role of GRB jets in driving the associated SNe, the fact that some SNe harbour jets brings up another natural question: are there choked relativistic jets in SNe that are not associated with GRBs, either of type Ic-BL or of other types. And if there are, what is the observational imprint of such jets. Recently it was argued that signatures of buried jets may have already been observed in stripped envelope SNe of various types (Ic-BL, Ic, Ib and IIb), which are not associated with GRBs  \citep{piran2019}. The point of this identification is that jets are expected to alter the velocity distribution of the SN ejecta so there is more energy carried by fast material compared to the expectation from a spherical explosion. The origin of this excessive fast material is the cocoon inflated by the jet within the stellar envelope, and can be seen either by emission \citep{ramirez-ruiz2002,nakar2017,de-colle2018,de-colle2018a} or by absorption \citep{piran2019,izzo2019,nakar2019}. In both cases the observations must take place at early times (typically within days of the explosion or even earlier).

In this context, low-luminosity GRBs ({\it ll}GRBs) are of special interest. These events are associated with Ic-BL SNe, which are very similar (in all aspects, including energetic, mass and composition) to those associated with regular LGRBs. But, the gamma-ray emission that accompanies the collapse of the core in these events is less luminous by about 4 orders of magnitude (and sometimes more) than the prompt gamma-rays of typical LGRBs. There are compelling evidence that the gamma-rays in \llgrbs  are not coming from the jet directly as in LGRBs. Instead it is suggested that the gamma-rays are emitted during a shock breakout \citep[e.g.,][]{kulkarni1998,campana2006,nakar2012}. In this scenario the breakout must be mildly relativistic and it takes place from a low-mass extended envelope \citep{nakar2015}. The most likely origin of the shock is the cocoon and it is expected to break out and generate gamma-rays also in case that the jet is choked within the extended envelope or even if it is choked within the compact core. If correct, then jets in \llgrbs are similar to those in LGRBs, with the main difference between the events being the structure of the progenitor envelope \citep{nakar2015}.

The general idea that jets may be involved in  driving SN explosions is old and was discussed by many authors \cite[e.g.][]{leblanc1970,bisnovatyi-kogan1970,ostriker1971,khokhlov1999,Burrows2007,Papish2011,Gilkis2014}. In the context of explosions that are driven entirely by GRB-like relativistic jets there have been only a few papers, all of which used numerical simulations where the jet is injected by hand at the core of the star. \cite{tominaga2007} explored the nucleosynthesis of a jet driven explosion, focusing on the composition of the outflow. \cite{lazzati2012} studied the dynamics of a jet driven explosion numerically. They found that a narrow jet that crosses the stellar envelope and successfully breaks out, such as in LGRBs, can deposit enough energy during its propagation to unbind the entire stellar envelope. The same is true for jets that are choked and die during their propagation. \cite{lazzati2012} found that in their simulations the slow bulk of the ejecta is accompanied by faster material, which can be mildly and/or ultra-relativistic, depending on whether the jet is successful and if not, on where in the envelope it is choked. \cite{barnes2018} and \cite{shankar2021} carried out simulations of SNe driven by relativistic jets. They studied the dynamics of the jet and the envelope, and focused on the light curves and spectra of the resulting SNe. They compared their numerically calculated light curves and spectra to SN Ic-BL observations, and deduced that a relativistic jet can simultaneously explode the star and generate the observed SN as well as break out of the progenitor and generate a LGRB. They concluded that there is no need of any additional source of energy at the center of the progenitor, apart from the jet, in order to explain the observations of SN/LGRB. 

There were also a couple of authors that considered a two-component explosion that contains a spherical source of energy at the core as well as a relativistic jet. \cite{suzuki2021} considered the effect of the jet on the explosive nucleosynthesis and on the chemical mixing  of newly synthesized heavy elements into high velocity outer layers of the ejecta. They found that the jet deposits a large amount of heavy metals in the high velocity layers, which is compatible with observations of Ic-BL SNe. Finally, \cite{de-colle2021} explored qualitatively various aspects of the interaction between a jet and a spherical (SN) explosion.

The goal of this paper is to study the possible range of effects that relativistic jets can have on the velocity distribution of SNe ejecta and the observational imprints that this distribution entails. We focus on the velocity distribution for several reasons. First, the velocity distribution can be probed via a variety of observations. It was done in a fair number of SNe in the past (both associated and unassociated with GRBs), and it is expected to be done on a regular basis in the near future with the increasing number of high cadence SN surveys. Second, it was shown that relativistic jets can have a significant, and possibly unique, effect on the velocity distribution of SN outflow \citep{lazzati2012}. Finally, Jets can leave their imprint on the velocity distribution even when they are choked, making it a unique tool to identify hidden jets. Most of the attention in this paper is given to successful jets. Choked jets are discussed only briefly and we present only a few examples of such jets. We defer a detailed discussion of choked jets to a future study.
 
After finding the velocity distributions that the jets generate, we use our results to ask whether narrow GRB relativistic jets can be the only energy source of the accompanied SN ejecta. Contrary to \cite{barnes2018} our answer is negative, as we find several different observations that rule out this possibility. We then explore whether a two component explosion, which includes a compact engine that launches a relativistic jet and a simultaneous release of energy that drives a quasi-spherical sub-relativistic explosion, can explain the velocity distributions inferred from the CCSNe discussed by \cite{piran2019}. Here we find a good agreement with observations. 

The paper is organised as follows. In \S\ref{sec:theory} we give an analytic description of the ejecta velocity distribution expected from an explosion driven by a relativistic jet. In \S\ref{sec:numerical} we report on a set of relativistic hydrodynamic simulations of jet driven explosions, and study the resulting velocity distribution of the ejecta. We also conduct a simulation of a two components explosion where the jet is launched simultaneously with an isotropic energy release at the center of the star. In \S\ref{sec:magentic} we discuss how jet magnetization is expected to affect the outflow velocity distribution. The observational signatures of the velocity distribution of jet driven explosions that were found in the previous section are discussed in \ref{sec:signatures}. We compare the results to available observations of GRBs and SNe in \S\ref{sec:observations}. We discuss previous studies in \S\ref{sec:previous}, and summarize our conclusions in \S\ref{sec:conclusions}.

\section{Outflow from a jet driven explosion - analytic considerations}\label{sec:theory}
We focus on the velocity distribution that relativistic jets induce on exploding stars, either when the jet is the source of the explosions or when it accompanies a spherical explosion. As we explain here, this distribution must be explored numerically since it involves mixing processes that cannot be followed analytically. Yet, analytic considerations can teach us about the processes that dominate the final distribution as well as the characteristic scales of the problem. Here we provide, first, an analytic approximation of the cocoon hydrodynamic evolution within the stellar envelope, which is the source of the final velocity distribution and is also of interest by itself. We then discuss the relevant scales that this evolution dictates for the outflow velocity distribution. 

\subsection{Successful jets}

\begin{figure*}
	\center
	\includegraphics[width=\textwidth]{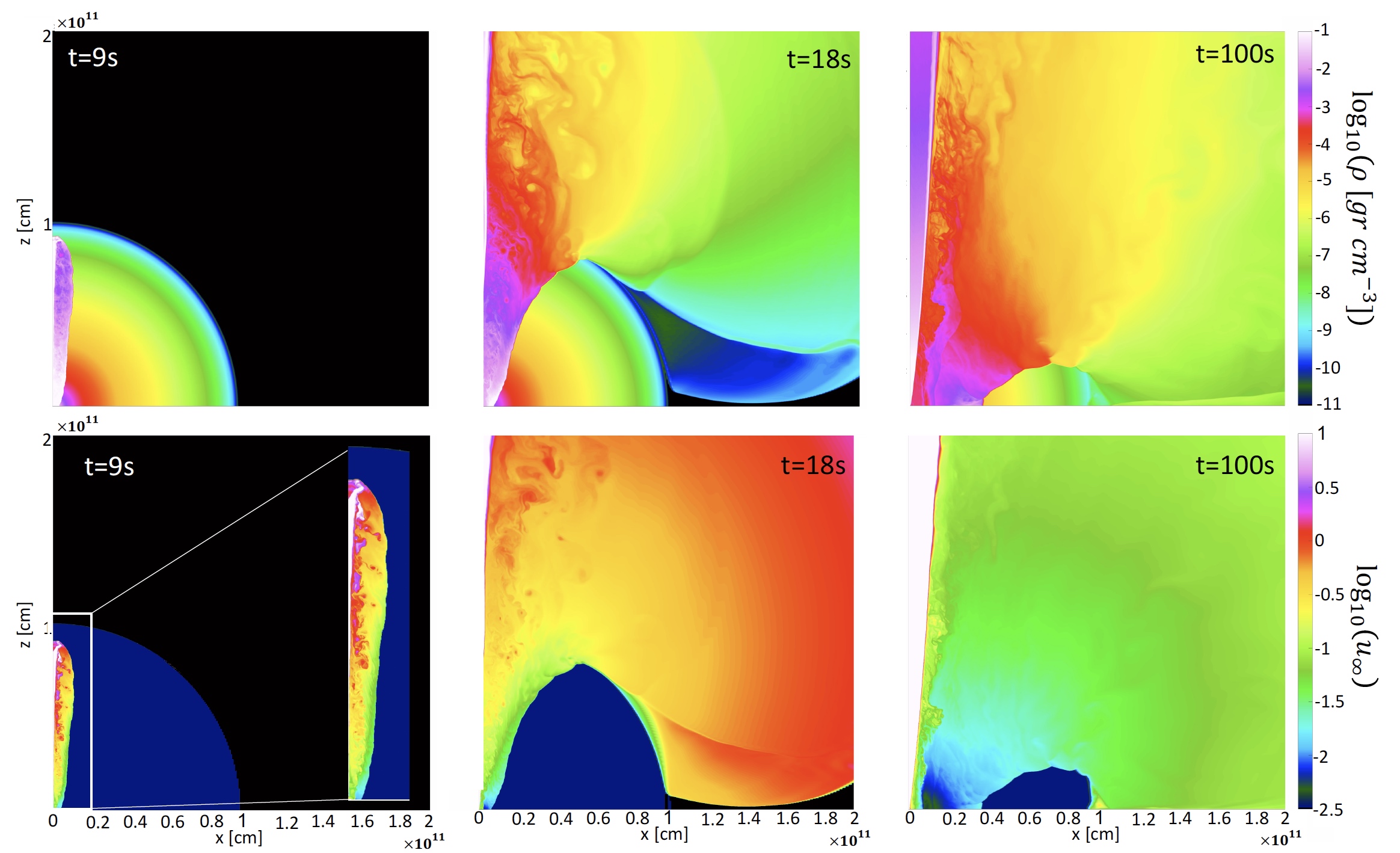}
	\caption{Colormaps of the density (top) and $u_\infty$ (bottom) from snapshots of three different times in a simulation of a successful jet exploding a star.  $u_\infty=\sqrt{\gamma^2h^2-1}$ ($h$ is the specific enthalpy) is comparable to the final proper velocity of a freely expanding fluid element, and to its energy per baryon in units of $m_p c^2$. The simulation parameters are $L_j=2 \times 10^{50}$erg/s, $\theta_j=0.1$rad, $M_\star=10{\rm M_\odot}$ and $R_\star=10^{11}$cm. The density profile is $\rho \propto r^{-2.5}(R_\star-r)^3$. The breakout time is about $10$ s so $E_c \approx 2 \times 10^{51}$erg and $v_0 \approx 0.01$c. In the bottom left panel one can see the distribution of $u_\infty$ in the inner cocoon, which is a direct result of the mixing. Near the head $u_\infty \sim 1-3$ (red to purple) while near the base $u_\infty \sim 0.1-0.3$ (green to yellow). In the bottom middle panel one can see that the velocity of the shock driven by the outer cocoon into the stellar envelope just after breakout is $\approx v_0/\theta_j \approx 0.1c$ (green color just behind the shock). In the bottom left panel it is shown that the shock velocity at the final stages, just before it consumes the entire envelope, is $\approx v_0 \approx 0.01c$ (pale blue color just behind the shock)
	}
	\label{fig1}
\end{figure*}

An analytic solution of the dynamics of the jet and the cocoon evolution during the jet propagation is derived by \cite{bromberg2011a} and calibrated numerically by \cite{harrison2018} (additional studies of jet propagation in dense media are found in, e.g., \citealt{blandford1974,marti1995,marti1997,begelman1989,aloy2000,macfadyen1999,macfadyen2001,matzner2003,zhang2004,lazzati2005,mizuta2006,mizuta2009,mizuta2013,morsony2007,morsony2010,lopez2013,lopez2016,gottlieb2019,gottlieb2020a,gottlieb2021c}). First we summarize the cocoon properties upon breakout, which set the initial conditions for the problem we solve here. 

The total energy deposited by the jet in the cocoon is the sum over the energy carried by jet material that crossed the reverse shock at the jet head until it breaks out:
 \begin{align}\label{eq:Ec}
     E_c &=\int_0^{t_b-R_\star/c} L_j dt \approx 10^{51} {\rm~erg} \left(\frac{L_j}{10^{50}{\rm~erg~s^{-1}}}\right)^\frac{2}{3} \times \notag\\
     & \times  \left(\frac{\theta_j}{0.1{\rm~rad}}\right)^\frac{4}{3}
	  \left(\frac{R_\star}{10^{11}{\rm~cm}}\right)^\frac{2}{3}\left(\frac{M_\star}{10{\rm~M_\odot}}\right)^\frac{1}{3}~~~;~~~ \rm{successful},
 \end{align}
where $t=0$ is the time that the jet launching starts and  $t_b$ is the breakout time (the time it takes the jet head to cross the stellar envelope). $L_j$ and $\theta_j$ are the jet luminosity and opening angle (upon launching), respectively, and $R_\star$ and $M_\star$ are the stellar envelope radius and mass, respectively. The approximation in the second equality of this equation is valid for jets with sub-relativistic heads, which are expected in almost all LGRBs. After breakout, the jet has more or less a clear path from the launching site all the way out of the progenitor star, therefore only a negligible fraction of the energy launched into the jet at $t>t_b-R_\star/c$ is deposited into the cocoon.
At the breakout time, the cocoon opening angle is comparable to the jet opening angle at launch, $\theta_j$, and it is composed of two distinct components - inner and outer. The energies in these components are comparable, so each carries about half of the total cocoon energy. A map of the terminal proper velocity $u_\infty $ (comparable to the proper velocity that corresponds to the energy per baryon) in the various components of the cocoon can be seen in inset at the lower left panel of Fig. \ref{fig1}.   

The {\it inner cocoon} is made out of jet material that was shocked at the head and spilled sideways during the jet propagation. This material is mixed with some stellar envelope material, which dominates the mass content of the inner cocoon and determines its temperature. The mixing of the inner cocoon was explored in detail in unmagnetized and weakly magnetized jets, but not in magnetically dominated jets. Note, however, that it is plausible that the fields of magnetically dominated jets are dissipated soon after launching, so by the time that the jet material arrives to the head it becomes weakly magnetized (\citealt{bromberg2016,gottlieb2021b}; see \S\ref{sec:magentic}). In unmagnetized and weakly magnetized jets the mixing takes place at three locations: (i) The head - a significant fraction of the mixing takes place as the shocked jet material spills from the jet head into the cocoon. Numerical simulations find that the energy per baryon of this material corresponds to a Lorentz factor of $\sim 2-3$, regardless of the initial Lorentz factor of the jet. (ii) The interface between the inner cocoon and the outer cocoon - the mixing along this interface is a continuous process that proceeds during the entire propagation of the jet. The result is a vertical density gradient in the inner cocoon with highest density (and mixing) near the base of the jet and lowest density near the jet head (see Fig. \ref{fig1}). As explained above, near the head the energy per baryon corresponds to particles Lorentz factor $\sim 2-3$. The level of mixing near the base varies depending on the system properties. In general, slower jets undergo more mixing near their base since the inner cocoon material has more time to mix before breakout. Numerical simulations find that for typical GRB parameters, where $t_b \approx 10$s, the energy per baryon of the inner cocoon near the base corresponds to particles velocity of $\sim 0.1$c.  (iii) The interface between the jet and the inner cocoon - mixing along this interface is facilitated by various types of instabilities and it depends on the jet and the medium parameters, most sensitively on the jet magnetization \citep{gottlieb2020,gottlieb2021}. This mixing can have a strong effect on the jet, but its effect on the cocoon, and thus on the Newtonian to mildly relativistic outflow is negligible. 

After breakout, the inner cocoon material starts escaping from the star through the cylindrical opening that the cocoon itself punched thorough the stellar surface. It is accelerating on the way out of the star as it converts its internal energy to a bulk motion.  The stratified mixing pattern of the inner cocoon, where more mixed colder material is at the bottom and less mixed hotter material is at the top, allows the hot material to escape first and the cold material to escape last so the final velocity of each fluid element can be approximated by its thermal velocity at the time of the breakout. The matter near the top, which escapes first, is accelerated to a Lorentz factor $\gamma \approx 2-3$. Its acceleration to mildly relativistic velocities prevents it from spreading to very large angles and it is expected to be confined to an opening angle $\sim 0.5-1$ rad around the jet axis. Slightly slower material breaks out after that, spreading to a slightly larger angle and so on until the slowest material that started near the base of the jet escapes. The velocity of the slowest inner cocoon material depends on the velocity of the jet head during its propagation, where for typical GRB parameter its velocity (measured in speed of light units) is $\beta \sim 0.1$.  Thus the inner cocoon of typical GRBs generates an outflow with a large range proper velocities $\gamma\beta \sim 0.1$ -- $3$. The fastest material is confined to $\theta \lesssim 0.5$ rad and material with $\beta \lesssim 1/2$ expands roughly isotropically.

The {\it outer cocoon} is composed only of shocked stellar material. Thus, the initial conditions at the time of the breakout involves no mixing and are therefore simpler. To estimate the shock velocity and its evolution, we approximate the energy behind the shock (i.e., in causal contact with the shock) as being constant as the shock crosses the stellar envelope from the pole to the equator until it consumes the entire envelope. This is a reasonable approximation if most of the 
stellar envelope shocked material does not escape from the star during the entire process. To see that this is the case, consider a fluid element at $r \approx R_\star/2$, where most of the envelope mass is. Its velocity after it crosses the shock is comparable to the shock velocity, and its distance from the stellar edge is comparable to the shock distance at this location from the equator. Therefore even if its velocity was radial and it was free to escape from the star (which it is not), this fluid element would have reached the stellar surface roughly at the time that the shock reaches the equator. The second approximation is that the pressure in the cocoon is roughly constant, $p_c\approx E_c/V_c$, where $V_c$ is the cocoon volume. We verify numerically that indeed the pressure does not vary by more than an order of magnitude over most of the shock front. Under this approximation the shock velocity at location $(r,\theta)$ is $v_{sh}(r,\theta) \sim \sqrt{(p_c(r,\theta)/\rho(r))} \sim \sqrt{(E_c/\rho V_c)}$. The typical density (of most of the envelope mass) is $M_\star/V_\star$, where $V_\star$ is the star volume, implying that the typical shock velocity when the outer cocoon occupies a volume $V_c$ is roughly $\sqrt{(E_c/M_\star)\times (V_\star/V_c)}$. At the time that the jet breaks out $V_c \approx V_\star \theta_j^2$, and when the shock consumes that entire envelope $V_c \approx V_\star$. Thus, the shock driven by the cocoon starts at a velocity $\approx v_0/\theta_j$ and ends at a velocity $v_0$, where we define $v_0=\sqrt{E_c/M_\star}$ (see Fig. \ref{fig1}). The range of terminal velocities achieved by the shocked envelope (outer cocoon) as it expands is expected to be comparable to the range of shock velocities, namely $\sim v_0$ -- $v_0\theta_j^{-1}$.

For typical GRB parameters where $E_c \sim 10^{51}$ erg, $M_\star \sim 10 {\rm M_\odot}$ and $\theta_j \sim 0.1$ rad. This corresponds to $v_0 \sim 0.01$c and therefore the velocity range of the outer cocoon is $\beta \sim 0.01$--$0.1$. The inner cocoon material is expected to carry a similar amount of energy in the proper velocity range $\gamma\beta \sim 0.1$--$3$. Thus, the expectation is that the combination of the two components will generate an outflow with a final proper velocity range  $\gamma\beta \sim 0.01$--$3$. The exact energy distribution within this range cannot be solved analytically. However, since the jets deposits a comparable amount of energy in the inner and outer cocoons, the energy carried by material in the proper velocity range 0.01--0.1 is comparable to the energy carried by material in the proper velocity range 0.1--3.

In addition to the velocity range of the outflow, the above discussion also provides two timescales that are important in case that the jet is accompanied by a spherical explosion (see discussion in \citealt{de-colle2021}). The first is the time it takes the cocoon material that breaks out with the jet to wrap the entire star as it spread sideways (i.e, reach the equator), $t_{wrap}$. This timescale is important since at $t>t_{wrap}$ the stellar surface can no longer be observed, and if there are regions along the stellar surface where the shock driven by the spherical explosion breaks out before the cocoon, they cannot be seen. As the velocity of cocoon material that moves towards the equator is $\sim 0.5$c and it has to cover a distance of $\sim \pi R_\star/2$  on its way to the equator we obtain:
\begin{equation}
	t_{wrap} \approx t_b+\frac{3R_\star}{c}
\end{equation}
For example in Fig. \ref{fig1}, $t_b \sim 10$s and $3R*/c=10$s and as can be seen in the middle panels,  $t_{wrap} \approx 20$ s. The second interesting timescale is the time it takes the jet to explode the entire star in case that it is the only energy source. This time is simply
\begin{equation}
	t_{exp,j} \approx C_{exp}(\alpha) \frac{R_\star}{\sqrt{E_c/M_\star}} 
\end{equation}
where $\alpha$ is a power-law index of the stellar envelope density, $\rho \propto r^{-\alpha}$. For example in Fig. \ref{fig1},  the entire enveloped is shocked roughly at 100 s while $\frac{R_\star}{\sqrt{E_c/M_\star}} \approx 300$ s, so $C_{exp}(\alpha=2.5) \approx 1/3$. Lower values of $\alpha$ results in higher values of $C_{exp}$, since the shock propagates mostly in the polar direction and smaller $\alpha$ implies high density at large radii. In our simulations we obtain $C_{exp}(\alpha=1) \approx 2/3$.

\subsection{Choked jet}
In this paper we only provide a general description of the outflow driven by collimated choked jets. A detailed study is deferred to a future work. The total energy in the cocoon of a chocked jet is simply the entire energy of the jet,
 \begin{equation}\label{eq:Ec_choked}
     E_c =\int_0^{t_e} L_j dt  ~~~;~~~ {\rm choked},
 \end{equation}
where $t_e$ is the total engine work time. The outflow from a choked jet depends mostly on the depth at which the jet is choked. A jet that is choked just before the breakout produces an outflow that is similar to that of a successful jet, without the jet component. Thus, the difference is apparent only within a narrow angle around the jet axis and only at $\gamma\beta \gtrsim 3$, where there is a cut-off in the outflow velocity. A jet that is choked more deeply generates an outflow with a cut-off at a lower velocity. If the jet is choked deep enough (according to our simulations at $r$ significantly smaller than $R_\star/3$), the inner cocoon component is also expected to be absent from the outflow. The reason is that as long as the jet is propagating, there is a fresh supply of energetic hot material into the inner cocoon. The supply of energy stops when the jet is choked, and the ongoing mixing of the inner cocoon as well as the continuous transfer of its energy to the freshly shock stellar material, reduces the inner cocoon contribution to the outflow until it becomes negligible. Finally, after the jet is choked, the cocoon becomes less and less collimated as it propagates, so its opening angle at the time of breakout is much larger than $\theta_j$ \citep{irwin2019}. Thus, for a deeply choked jet, the maximal velocity of the outflow has a a maximal velocity that is lower than $v_0/\theta_j$. This maximal velocity drops with the depth at which the jet is choked.

\section{Numerical simulations}\label{sec:numerical}
We perform a set of numerical simulations to explore the effect that various jet and progenitor properties have on the velocity distribution of the outflow. We use mostly 2D simulations since they demand less computational resources and allow for higher resolution. We expect 2D to be accurate since the difference between 3D and 2D simulations lies mostly in the instabilities that develop along the jet-cocoon interface and in the shape of the jet head (both of which have a minor effect on the cocoon), whereas the mixing at the head and at the interface between the inner and outer cocoon is similar \citep{harrison2018}. We verify that by running a 3D simulation and comparing it to the result of a 2D simulation (see appendix \ref{sec:3D}). Below we describe first the setup of our simulations, followed by a discussion of their results. 

\subsection{Simulations setup}
We carry out relativistic hydrodynamic (RHD) numerical simulations using \textsc{pluto} version 4.0, a freely distributed numerical code \citep{mignone2007}. We use the \textsc{pluto} special RHD module to perform a large set of 2D simulations with a Cylindrical coordinate system, and one 3D simulation with a Cartesian coordinate system, both with a static grid. We inject the jet by hand as a boundary condition at the base of the grid using the injection method of \cite{mizuta2013}. Specifically, we inject a relativistic hot cylindrical jet along the $\hat{z}$-axis through a nozzle at a height $Z_i$ and a cylindrical radius $r_i$, with a bulk Lorentz factor $\Gamma_0$, for a limited time $t_e$. 
The injected flow spreads sideways relativistically forming a jet with a roughly uniform luminosity $L_j$ over an opening angle of $\theta_j=\frac{1}{f\Gamma_j}$, where $f=1.4$ is a constant \citep{harrison2018}. In most simulations $Z_i= 1 \times 10^{9} \,\mathrm{cm}$, $r_i=8 \times 10^{7} \,\mathrm{cm}$ and $\Gamma_0=5.0$. In two cases the injection is made at $Z_i=0.67 \times 10^{9}~[2 \times 10^9] \,\mathrm{cm}$ and with $\Gamma_0=3.33 [10]$, so the opening angle of the jet is different by a factor of 1.5 [0.5] compared to our canonical case.
For the progenitor star, we take a constant radius $R_\star= 10^{11}\,\mathrm{cm}$ and a density profile $\rho_\star=Ar^{\alpha}(R_\star-r)^{n}$, where $\alpha$ and $n$ are constants and $A$ is a constant chosen so the stellar mass is $10M_\odot$. Simulations that include a spherical explosion were made by planting the explosion energy as initial uniform pressure at $r < 10^9\,\mathrm{cm}$, at the beginning of the simulation. In the simulation that includes both a spherical component and a jet we inject the jet at $Z_i=1 \times 10^{8} \,\mathrm{cm}$ and $r_i=8 \times 10^{6} \,\mathrm{cm}$.
Throughout the simulations we apply an ideal equation of state with a constant adiabatic index of $4/3$, as appropriate for a radiation dominated gas. We neglect gravity forces in our simulations. This is justified since the escape velocity from the stellar surface, $0.005c$, is much lower than the values of $v_0$ in all our simulations. Thus, gravity would have a negligible effect on elements that are ejected at $v>v_0$, and the timescales involved in the jet driven explosion are much shorter than those of the progenitor star’s collapse.

We vary the jet and progenitor parameters $t_e, L_j, \theta_j, \alpha$ and $n$. In all our jet driven explosion simulations the jet is successful, except for two simulations. In one the jet is choked after crossing about 30\% of the stellar envelope, and in another the jet launching stops when the head is close to the stellar surface so the last bit of the launched jet is choked just as it reaches the stellar surface. We also carry out one simulation with a purely spherical explosion and one simulation with a combined jet and a spherical explosion. The properties of all the simulations we perform are given in Table~\ref{simulations_configurations}. In order to verify that we can use 2D simulations to study the velocity distribution we use two simulations with an identical physical configuration, one in 3D and one in 2D. These two simulations are presented as model $Lc$ in \citet[][]{gottlieb2021}, where their numerical setup can be found (see discussion in appendix \ref{sec:3D}).

As a canonical grid, we take 3000 grid points in the $ \hat{z} $-direction, 1000 points are uniformly distributed from $z_i$ to $z=10^{11}\,\mathrm{cm}$ (one stellar radius), and the rest of them have logarithmic mesh spacing at larger distances, from $z=10^{11}\,\mathrm{cm}$ to $z=10^{13}\,\mathrm{cm}$ ($100$ stellar radii). In the $ \hat{r} $-direction we take 2200 grid points, 200 points are uniformly distributed from $r=0$ to $r=1 \times 10^{9} \,\mathrm{cm}$, and the rest of them have logarithmic mesh spacing at larger distances, from $r=1 \times 10^{9}\,\mathrm{cm}$ to $r=10^{13}\,\mathrm{cm}$.
Simulations that include a spherical explosion have additional 100 more grid points in the $ \hat{r} $-direction, that are uniformly distributed from $r=0$ to $r=5 \times 10^{7} \,\mathrm{cm}$, in order to keep high resolution at the jet injection nozzle. The next 200 points have logarithmic mesh spacing from $r=5 \times 10^{7} \,\mathrm{cm}$ to $r=1 \times 10^{9} \,\mathrm{cm}$, and the last 2000 points remain the same. Our convergence test is described in appendix \ref{app:convergence}.
We stop the simulations long after the entire envelope is shocked, and verify that all the material (except for some that is much slower than $v_0$) have reached its final velocity.

\begin{table*}[t]
   
     \setlength{\tabcolsep}{8.0pt}

                \centering

                \begin{tabular}{ | l | c  c  c  c  c  c | }

                    \hline

                    Jets& $ L_j [10^{51}\rm{erg~s^{-1}}] $ & $ \theta_{j}[deg] $ & $ u_{\infty,max}$ & $ \rho_\star(r) $ & $ t_b [\rm{s}] $ & $ t_e [\rm{s}] $\\ \hline
                    
                    Canonical & $ 1.0 $ & $ 8 $ & $ 500 $ & $ \propto r^{-2}x^3 $ & $ 8.4 $ & $ 30 $\\
                    
                    $\alpha 2.5$ & $ 1.0 $ & $ 8 $ & $ 500 $ & $ \propto r^{-2.5}x^3 $ & $ 6.9 $ & $ 8.33 $\\

                    $\alpha 2.8$ & $ 1.0 $ & $ 8 $ & $ 500 $ & $ \propto r^{-2.8}x^3 $ & $ 4.5 $ & $ 8.33 $\\  
                    
                   $n1$ & $ 1.0 $ & $ 8 $ & $ 500 $ & $ \propto r^{-2}x^1 $ & $ 8.4 $ & $ 8.33 $\\   
                    
                    $\theta 4$ & $ 1.0 $ & $ 4 $ & $ 1000 $ & $ \propto r^{-2}x^3 $ & $ 4.5 $ & $ 8.33 $\\
                                        
                    $\theta 12$ & $ 1.0 $ & $ 12 $ & $ 333.3 $ & $ \propto r^{-2}x^3 $ & $ 13.5 $ & $ 30 $\\
                    
                    Strong & $ 3.16 $ & $ 8 $ & $ 500 $ & $ \propto r^{-2}x^3 $ & $ 6.9 $ & $ 18 $\\
                    
                    Weak & $ 0.1 $ & $ 8 $ & $ 500 $ & $ \propto r^{-2}x^3 $ & $ 18.8 $ & $ 30 $\\                    

                    Barely Choked & $ 1.0 $ & $ 8 $ & $ 500 $ & $ \propto r^{-2}x^3 $ & $ - $ & $ 6.2 $\\

                    Choked & $ 1.0 $ & $ 8 $ & $ 500 $ & $ \propto r^{-2}x^3 $ & $ - $ & $ 2.55 $\\ \hline \hline


                    Spherical & $ L_j [10^{51}\rm{erg~s^{-1}}] $ & $ \theta_{j}[deg] $ & $ u_{\infty,max}$& $ \rho_\star(r) $ & $ E_{sph} [10^{51}\rm{erg}] $ & $ t_e [\rm{s}] $ \\ \hline

                    & $ - $ & $ - $ & $ - $ & $ \propto r^{-2}x^3 $ & $ 5.6 $  & $ - $\\\hline \hline

                    Jet + Spherical & $ L_j [10^{51}\rm{erg~s^{-1}}] $ & $ \theta_{j}[deg] $ & $ u_{\infty,max}$ & $ \rho_\star(r) $ & $ E_{sph} [10^{51}\rm{erg}] $ & $ t_e [\rm{s}] $ \\ \hline

                    & $ 1.0 $ & $ 8 $ & $ 500 $ & $ \propto r^{-2}x^3 $ & $ 41.7 $  & $ 8.33 $\\\hline

                \end{tabular}

                \hfill\break

                \caption{The simulations' parameters. $ L_j $ is the jet luminosity (two sided), $ \theta_{j} $ is the jet (half) opening angle upon launching  (in degrees), $ u_{\infty,max} = \sqrt{\Gamma_0^2h_0^2-1} $ is the terminal four-velocity of the jet, and $ \rho_\star(r) $ is the density profile of the star, where $r$ is the distance from the center and $ x \equiv (R_\star-r)$ is the distance from the stellar edge. In all simulations the stellar mass is $10M_\odot$ and the stellar radius is $R_\star=10^{11}$ cm. $ t_b $ is the jet breakout time and $ t_e $ is the central engine duration.}\label{tab_models_comparison}

         \label{simulations_configurations}

\end{table*}

\subsection{Results}
\subsubsection{Jet driven explosions}
\begin{figure}
	\center
	\includegraphics[width=0.45\textwidth]{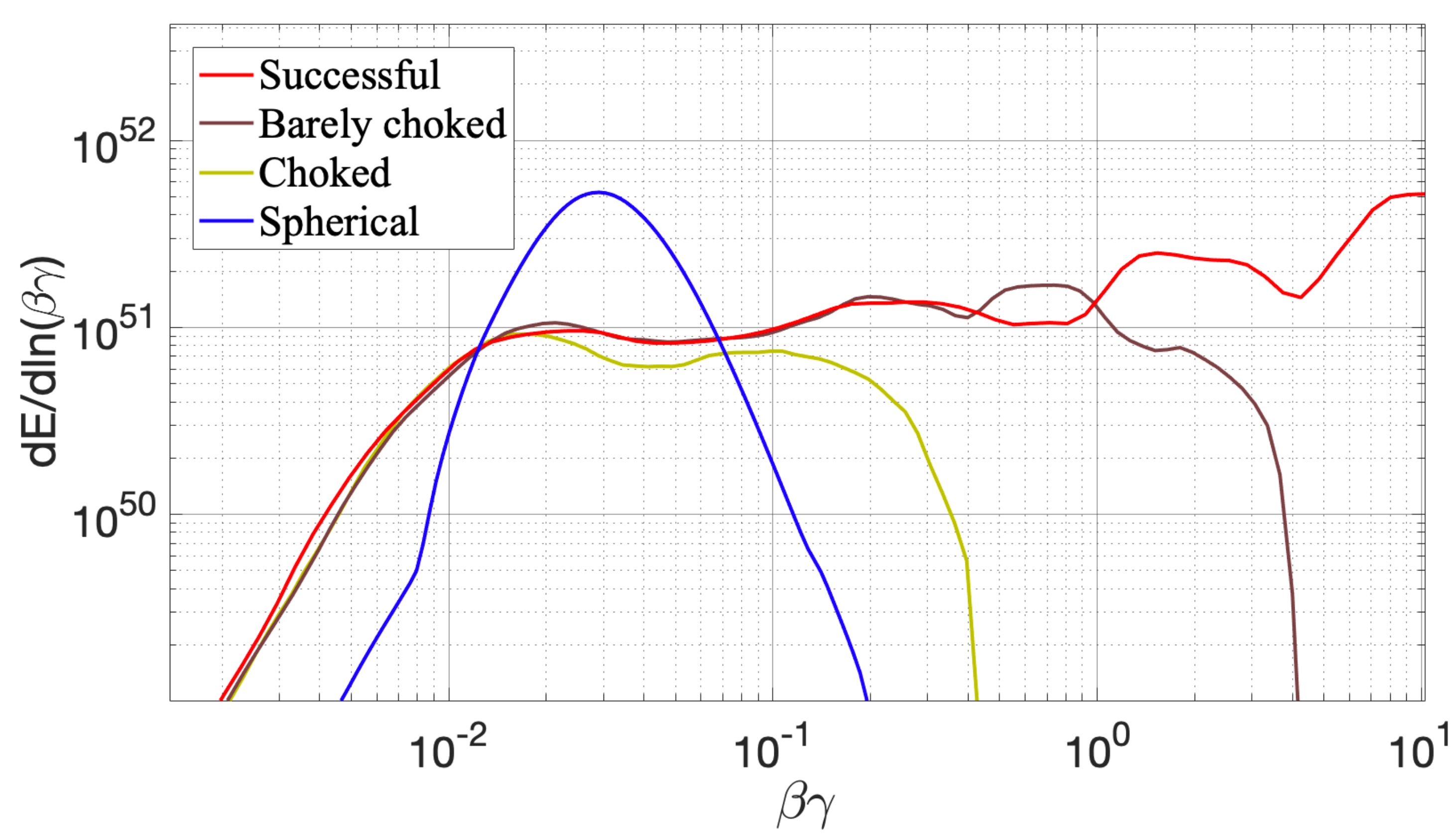}
	\caption{The energy distribution as a function of the proper velocity of three jet driven stellar explosions. In all simulations the progenitor and jet properties are the same, except for the engine work time. In one $t_e \gg t_b$ (successful jet), in the second the jet launching stops so it is choked just upon breakout (barely choked), and in the last the jet is choked after reaching approximately 30\% of the stellar radius (choked). In the successful and barely choked jets $v_0=0.017$c, and in the choked jet $v_0=0.012$c. For comparison we also include the energy-velocity distribution of a spherical explosion of the same progenitor with energy that is similar to $E_c$ in the successful and barley choked cases. In the spherical explosion all the energy is deposited in the bulk of the mass (unlike a jet driven explosion where only the outer cocoon energy is deposited into the envelope), and therefore for the spherical explosion $v_0=\sqrt{2E/M_\star}=0.025$c.
	}
	\label{fig:canonical}
\end{figure}

Fig. \ref{fig:canonical} shows distributions of the energy per  logarithmic scale of the proper velocity, $\frac{dE}{d\ln(\gamma\beta)}$, of three jet driven explosions that have reached the homologous phase. The jet and progenitor properties are typical for LGRBs. Note that the relation between this distribution and the density distribution, which is typically plotted in papers about SNe, satisfies in the sub-relativistic regime $\rho(v) \propto \frac{dE}{d\ln(v)} v^{-5}$. All simulations show outflows with a roughly constant amount of energy per logarithmic scale of proper velocity over a velocity range that starts at about $v_0$ and ends at a velocity that depends on whether the jet is choked or not. If the jet is successful, the distribution continues to ultra-relativistic velocities, and if it is barely choked, the ultra-relativistic component of the jet is absent and the distribution includes only the cocoon. As expected, the distribution of the cocoon spans the range $\gamma\beta \approx \beta_0$--$3$, where $ \beta_0 = v_0/c$. The (deeply) choked jet explosion has no relativistic component and the outflow velocity range is  $\beta \approx \beta_0$--$0.2$. For comparison we also show the distribution of a spherical explosion, where the energy is tightly concentrated around the characteristic velocity, $v_0$. 
\begin{figure}
	\center
	\includegraphics[width=0.47\textwidth]{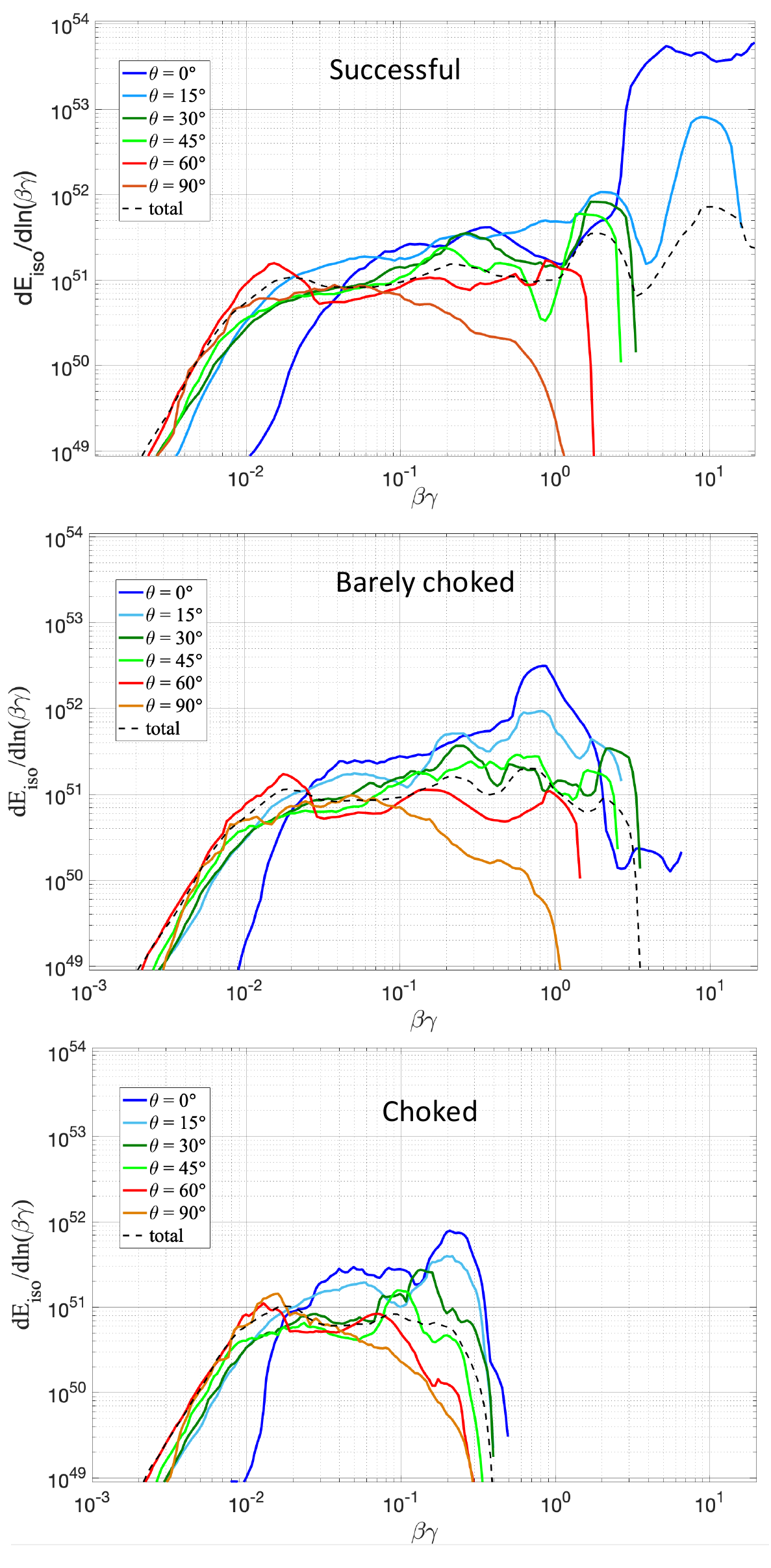}
	\caption{The {\it isotropic} energy distribution as a function of the proper velocity, $\frac{4\pi dE(\theta,\beta\gamma)}{d\Omega d\ln(\beta\gamma)}$, in different directions compared to the jet axis in the same simulations shown in Fig. \ref{fig:canonical}.
	}
	\label{fig:angular}
\end{figure}

Fig. \ref{fig:angular} shows the angular dependence of the outflow energy distribution for the same three simulations shown in Fig. \ref{fig:canonical}. As expected, when the jet is successful (top panel), the entire ultra-relativistic component is concentrated around the jet axis at an angle $\lesssim 15^\circ$, and is therefore part of the jet, which is not of interest in this paper\footnote{Note that the energy distribution and angular structure of the jet cannot be properly simulated in 2D (see discussion in \citealt{gottlieb2021}), and therefore the only thing that can be taken from the distribution of the ultra-relativistic outflow in our 2D simulations  is that it is concentrated in a narrow angle around the jet axis.} (for a detailed discussion of the structure of the jet in various scenarios see \citealt{gottlieb2020a,gottlieb2020,gottlieb2021,gottlieb2021a}). As expected, the mildly relativistic outflow with $\gamma \approx 2$--$3$ is limited to within a polar angle of $\sim 30-45^\circ$ and material with $\gamma\beta \approx 1$ is limited to $\sim 60^\circ$. The source of all this material with $\beta \gtrsim 0.1$ is the inner cocoon. Most of the material from the outer cocoon is expanding at $\theta \gtrsim 60^\circ$ and it dominates the outflow energy in the range 0.01c--0.1c. The barely choked jet (Fig. \ref{fig:angular}, middle panel) is similar to the successful jet, except for the absence of the ultra-relativistic component. Thus, the distributions are almost identical at $\theta \gtrsim 30^\circ$. In fact, the barely choked jet shows, roughly,  the contribution of the cocoon to the jet-cocoon outflow in the case of a successful jet. Finally, the deeply choked jet (Fig.  \ref{fig:angular}, bottom panel) shows no material faster than $0.3$c, and it is rather similar to the successful and barely choked jets at velocities lower than about 0.3c. The outflow with $0.1c\lesssim v\lesssim0.3c$ is composed out of inner cocoon material that was mixed heavily but still survived until the breakout. It is therefore confined mostly to $\theta \lesssim 30^\circ$. We expect that jets which are choked significantly deeper than $0.3R_\star$ would not have this component at all and their maximal outflow velocity in all directions will be significantly lower than $v_0/\theta_j$, owing to the complete mixing of the inner cocoon and the expansion of the outer cocoon opening angle before the breakout.

As evident from Figs. \ref{fig:canonical} and \ref{fig:angular}, the results of the specific simulations that are depicted in these figures fit our theoretical predictions regarding the range of velocities over which the outflow spreads and its angular distribution. We are therefore quite confident that the theoretical predictions are also valid for other jet parameters. However, theory cannot predict the exact velocity distribution, except for the fact that the energy of the inner and outer cocoons in a successful jet are comparable. Thus, the fact that the energy distribution is flat (in log space) over the entire proper velocity range must be tested over a range of parameters to see if it is general or not. We therefore carry out a set of simulations of successful jets varying the two jet parameters, $L_j$ and $\theta_j$, and the two stellar structure parameters, $\alpha$ and $n$. The mass of the star is not an independent parameter, since the jet evolution depends on $L_j/\rho$ and thus reducing $L_j$ by some factor is similar to increasing $M_\star$ by the same factor. Finally, $R$ may have an effect, apart for its effect on $\rho$ which is explored by varying $L$, but only if $\alpha \neq 2$, since for $\alpha=2$ the jet structure is self similar and the head velocity is constant (see \citealt{harrison2018}). Since in stars $\alpha$ is not expected to be very different than 2, we do not vary $R$.

\begin{figure}
	\center
	\includegraphics[width=0.48\textwidth]{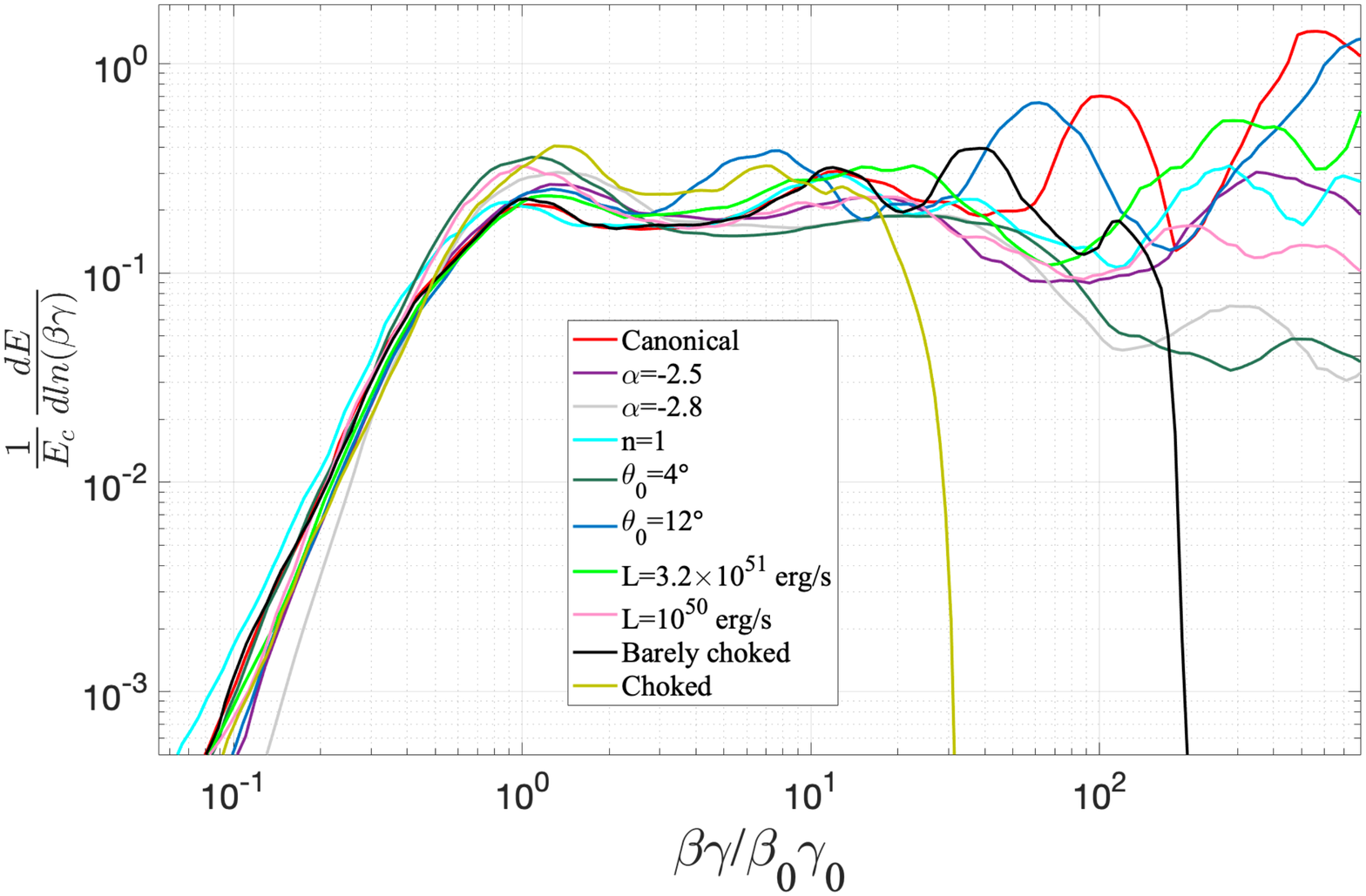}
	\caption{The energy distribution as a function of the proper velocity of all jet driven explosion simulations at the homologous phase. The energy of each explosion is normalized by $E_c$, and the proper velocity is normalized by $\beta_0\gamma_0$. This figure demonstrates that jet driven explosions exhibit a universal quasi-uniform energy distribution over each logarithmic scale of the outflow proper-velocity. The increased spread between the curves at $\gamma\beta > 100 \gamma_0\beta_0$ corresponds to $\gamma\beta \gtrsim 3$ and it reflects the structure of the jets, which are not of interest in this paper.
	}
	\label{fig:allJets}
\end{figure}

Fig. \ref{fig:allJets} shows the velocity profiles of all outflows in our jet driven explosion simulations. The energy of each explosion is normalized by $E_c$, and the proper velocity is normalized by $\beta_0\gamma_0$.
This figure shows that over a very wide range of parameters, that covers the relevant parameters for LGRBs, jet driven explosions generate outflows with a universal flat kinetic energy profile over a large range of proper velocities (more than an order of magnitude). The only difference between the simulations is the velocity range over which the flat profile is observed, which matches the theoretical expectation, and depends mostly on whether the jet is choked or not. 
As we discuss later this type of outflow is inconsistent with observations of GRBs and SNe.


\subsection{A two component explosion - jet and spherical}

\begin{figure}
	\center
	\includegraphics[width=0.45\textwidth]{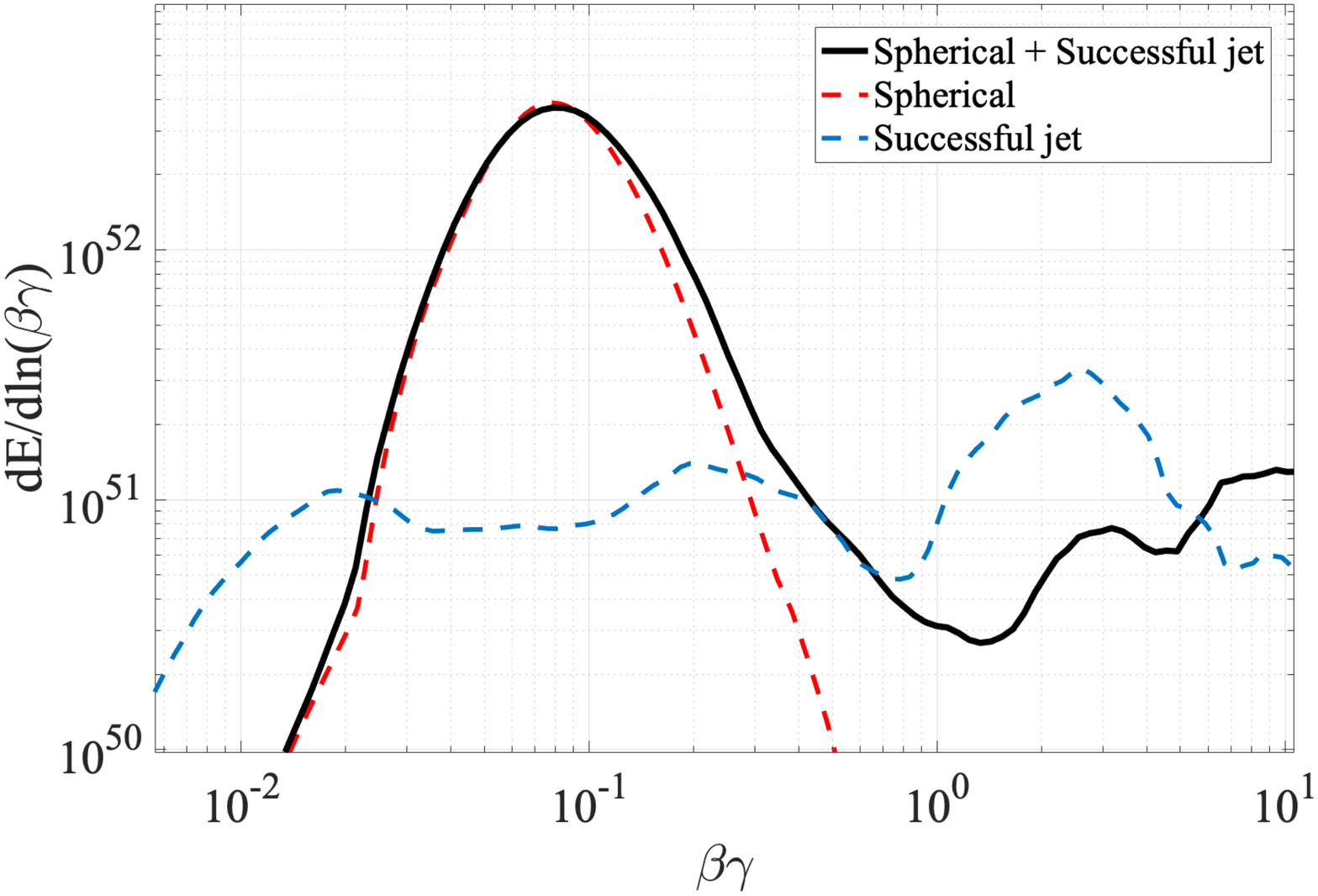}
	\includegraphics[width=0.45\textwidth]{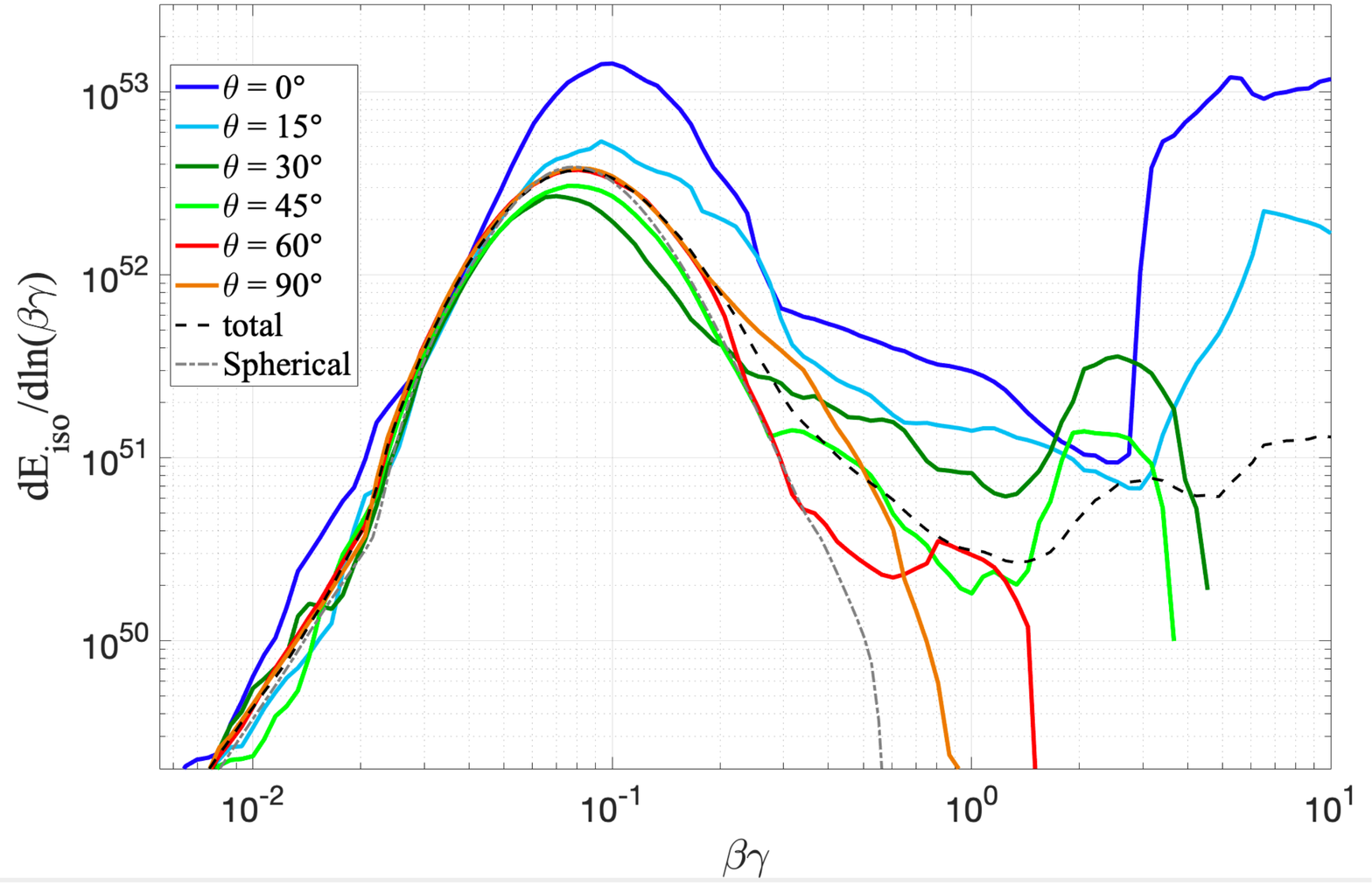}
	\caption{The energy distribution as a function of the proper velocity during the homologous phase of a two component explosion - spherical and jet.  The {\it top panel} shows the total distribution as well as the distributions of two single component explosions, one by a spherically symmetric energy release and another by a relativistic jet. The {\it bottom panel} shows the distribution of the isotropic energy of the outflow from the two-component explosion in different directions. For comparison we also include the total distributions of the two-component explosion and spherical explosion.} 
	\label{fig:Jet+SN}
\end{figure}

As we discuss below, there are many observed SNe and GRBs that cannot be explained neither by a spherical explosion nor by a jet. Our goal here is just to find out whether having both components in a single explosion, namely a jet that is accompanied by a spherical release of energy, can potentially produce an outflow with a velocity distribution that is compatible with the observations. Note that we explore only a single setup and we do not examine the wide range of possible interactions between jets and spherical explosions. The top panel of Fig. \ref{fig:Jet+SN} depicts the energy distribution as a function of the proper velocity of a combined explosion, where we launch the jet and release the spherical blast wave at the center of the progenitor simultaneously. Guided by GRB observations, the spherical explosion energy, $E_{sph}$ is significantly higher than the energy deposited by the jet in the cocoon, $E_c$ (see Tab. \ref{tab_models_comparison}). The top panel of Fig. \ref{fig:Jet+SN} also includes the outflow distribution from two other simulations, one with a similar jet alone and another with a similar spherical explosion without a jet. The figure shows that the outflow distribution from the simulation with the two energy sources is rather similar to a simple superposition of the outflows from the two simulations, each with a single energy sources. There are some differences though. First, the low velocity material in the jet driven explosion is absent, as it was driven to higher velocities by the more energetic spherical explosion. Second, the energy at mildly relativistic velocities in the two-component explosion is a bit lower than the energy in the jet only case. The reason is a slightly enhanced mixing in the inner cocoon of the two-component explosion. The stronger mixing causes a larger fraction of the inner cocoon material to expand at sub-relativistic velocities, thus reducing the energy of the part that expands to mildly relativistic velocities.
We do not explore the origin of this small difference in the mixing (and whether it is due to physical or numerical processes). Nevertheless, the picture is clear, when an energetic spherical explosion and a jet with less energy than the spherical component are combined,  the result is an energy distribution with a prominent peak at $\sim \sqrt{2E_{sph}/M_\star}$, and an additional high velocity component that can extend to mildly relativistic velocities whose energy is comparable to $E_c$. 

The bottom panel of Fig. \ref{fig:Jet+SN} depicts the angular distribution of the outflow from the two component explosion. It shows that, as expected, the fast velocity material that is associated with the jet and the cocoon is most prominent at $\theta \lesssim 60^\circ$. Yet, also along the equator there is a potentially observed access of high velocity outflow (compared to the spherical explosion). This figure shows that some of the observables depend on the viewing angle (e.g., spectral absorption feature).

\begin{figure}
	\center
	\includegraphics[width=0.5\textwidth]{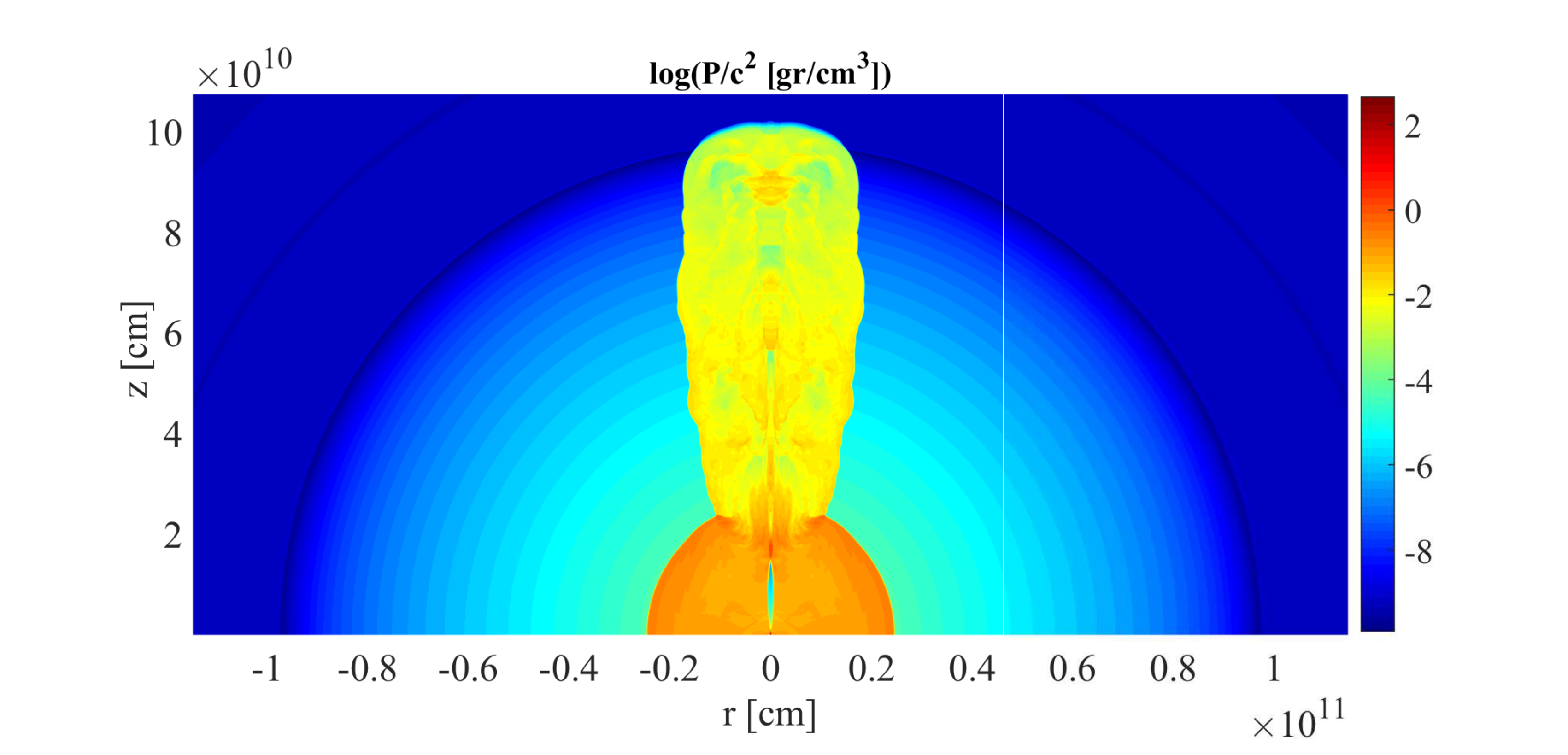}
	\caption{A colormap of the pressure from a simulation of a combined case including both a relativistic
jet and a spherical explosion, at the moment of the jet breakout at $ t = 8$ sec.}
	\label{fig:Jet+SN_map}
\end{figure}

The reason that the distribution of the combined explosion resembles a superposition can be understood from the pressure map of the combined explosion simulation at the moment of the jet breakout, as shown in Fig. \ref{fig:Jet+SN_map}. The jet head and the cocoon are breaking out of the stellar surface while the spherical explosion is still deep within the star. Thus, the fast cocoon material which escapes the progenitor is causally disconnected from the spherical explosion and the energy distribution of the fast moving material resembles that of a purely jet driven explosion . The spherical shock wave, which is driven by an explosion that is much more energetic than the jet, sweeps the slowly moving material and the unshocked regions of the envelope without being affected by the lower amount of energy deposited by the jet. Consequently, the velocity distribution of the slow moving material is similar to that of a spherical explosion.

Finally, while we do not carry out simulations of a spherical explosion that is combined with a choked jet, the outcome of such a scenario is expected to take the following form. Since the fast velocity material of the cocoon is not strongly affected by the spherical explosion, we expect that the effect of a choked jet will be similar to the case of a jet driven explosion. Namely, the deeper the choking takes place, the lower is the cutoff at high velocity. Thus, if $E_{sph}$ is significantly larger than $E_c=E_j$, and the choking takes place well above the location of the spherical blast wave at the choking time, then the final velocity distribution of the outflow will include an excess of fast moving material compared to a spherical explosion. If, however, the choking takes place deep enough, such that the spherical ejecta sweeps the cocoon, then it will erase most of the evidence to the jet existence, at least in the velocity profile that might be indistinguishable by observations from a spherical explosion without a jet.


\section{The effect of jet magnetization on the velocity distribution}\label{sec:magentic}
All the jets in our simulations are unmagnetized. However, it is likely that GRB jets are launched with a non-negligible magnetic field, possibly even Poynting flux dominated. It is therefore interesting to consider the effect of jet magnetization on the velocity distribution of the outflow. It is first important to note that the jet propagation is sensitive mostly to the magnetization level below the jet head and this magnetization is not necessarily the same as the one at the base of the jet. \cite{bromberg2016} used relativistic magnetohydrodynamic (RMHD) simulations to explore the propagation of jets that are dominated by magnetic fields at the launching site, i.e., with initial magnetization $\sigma_0 \gg 1$, where $\sigma=\frac{b^2}{4\pi h\rho c^2}$ and $b$ is the magnetic field in the fluid rest frame. They found that Poynting flux dominated jets in GRBs are collimated to form a nozzle where magnetic fields are prone to dissipation, facilitated by kink instability. The conditions for magnetic dissipation at the nozzle, as well as the final jet magnetization above the nozzle are not fully understood, but it is possible that in a wide range of conditions the magnetization in the jet drops at the nozzle to $\sigma \sim 1 $ or even lower. For example, recently \citet{gottlieb2021b} performed a general-relativistic magnetohydrodynamic (GRMHD) simulation of LGRB jets in stars, and found a similar phenomenon. However, in their models the dissipation was stronger, and the jet became weakly-magnetized with $ \sigma \ll 1 $.
Further studies are needed in order understand the outcome of the propagation of a Poynting flux dominated jet through the nozzle and its magnetization above the nozzle.

Magnetic fields in the jet (above the nozzle and below the head) may affect the outflow velocity distribution in two ways. First, magnetic fields can make the jet head narrower and consequently to propagate faster \citep[e.g.,][]{gottlieb2020}. Thus, a magnetized jet may deposit less energy into the cocoon than an unmagnetized jet with otherwise similar properties, leading to a lower $v_0$.
Second, a magnetized jet forms a magnetized inner cocoon. This can reduce the mixing in the inner cocoon, which in turn will affect the final velocity distribution of its material. This is probably not the case in weakly magnetized jet. \cite{gottlieb2020} studied the propagation of weakly magnetized  jets up to the point of breakout and they found that weak magnetization has no significant effect on the mixing of the inner cocoon, at least for $\sigma \lesssim 0.1$ (see, for example, their figure 1). The effect of $\sigma \gtrsim 1$ on the mixing in the inner cocoon has not been explored. It may be similar to an unmagnetized jet, or alternatively, if there is less mixing then the final velocity of the inner cocoon material will be mildly and/or ultra-relativistic. In any case, regardless of the jet magnetization, the inner cocoon is expected to contain about half of the total cocoon energy. We therefore conclude that while we cannot determine the precise final velocity distribution of the inner cocoon material in a case of a highly magnetized jet, it is expected to be in the mildly and possibly even ultra-relativistic range and carry about half of the outflow energy. Finally, one region that is not affected by the jet magnetization is the outer cocoon (apart for, possibly, a lower value of $E_c$). Thus, similarly to unmagnetized jets, the outer cocoon is expected to generate a flat energy distribution (in log-space) between $v_0$ and $v_0/\theta_j$, which contains about half of the outflow energy. 

To summarize, the velocity distribution of an explosion driven by a magnetized jet is expected to be similar in many aspects to that of an unmagnetized jet. It covers a large range of velocities, at least over a decade, with a flat energy profile (in logarithmic velocity space) in sub-relativistic velocities and a comparable energy at mildly to ultra-relativistic velocities. 

\section{Observational signatures of a jet driven explosion}\label{sec:signatures}
The velocity distribution of the outflow from a jet driven stellar explosion has a range of observational signatures, which we discuss below.

{\it The energy in the sub-relativistic, mildly relativistic and ultra-relativistic components}: Each of the outflow components dominates a different part of the emission. The sub-relativistic outflow dominates the SN emission. The ultra-relativistic ejecta dominates the LGRB and the afterglow on timescales of days to weeks (for observers within the opening angle of the jet). Finally, the mildly relativistic outflow dominates the late radio emission if it carries more energy than the ultra-relativistic component as well as the early emission for observers that are away from the beam of the ultra-relativistic jet. We therefore denote the energies of the different components as $E_{SN}$, $E_{MR}$ and $E_{LGRB}$. Our results show that in a jet driven explosion  $E_{SN} \sim E_{MR} \sim E_c/2$ (regardless of the jet magnetization), unless the jet is choked deep within the star. The ultra-relativistic energy, $E_{LGRB}$ can in principle be larger or smaller than $E_{SN}$. In a choked jet $E_{LGRB}=0$, while if the jet continues to work much longer than the breakout time, then  $E_{LGRB} \gg E_{SN}$. However, as we argue below in observed LGRB/SNe, if those are jet driven explosions, then typically we expect $E_{SN} \sim E_{LGRB}$.

During the propagation of the jet through the stellar envelope, the mixing of the jet material as it flows from the head to the cocoon implies that jet material that crosses the reverse shock will never re-accelerate to ultra-relativistic velocities. This property is shown clearly in the barely choked jet simulation. Thus, in case of a successful jet $E_c$ is given by Eq. \ref{eq:Ec} and $E_{LGRB} \approx \int_{t_b-R_\star/c}^{t_e} L_j dt$. The roughly constant luminosity of the prompt emission (when averaged over the rapid temporal fluctuations)\footnote{LGRBs are highly variable but the luminosity of the various pulses does not seems to vary, on average, during the entire duration of the burst. Namely, pulses that are seen at the beginning of the pulse are as bright, on average, as pulses seen towards the end of the prompt emission \citep[e.g.,][]{nakar2002}.} suggests that $L_j$ does not increase or decrease significantly over the GRB duration. Since the jet engine does not know about the breakout, it stands to reason that the average jet luminosity does not vary much also at $t<t_b$. Thus, we can approximate $E_c/E_{LGRB} \approx \frac{t_b}{t_e-t_b}$, where we approximate $t_b \gg R_\star/c$, as appropriate for a sub-relativistic jet head. Now, since the jet engine is not affected by the breakout, we obtain that only in a small fraction of successful jets $t_e-t_b \ll t_b $ (this fraction is quantified in \citealt{bromberg2013}). The duration distribution of GRBs indicates that the number of observed GRBs with $t_e-t_b \gg t_b $ is also small \citep{bromberg2013}. We therefore conclude that in typical LGRBs $t_e-t_b \sim t_b$ and thus if GRB/SNe are generated by jet driven explosions then $E_{LGRB} \sim E_c \sim E_{SN}$ (this conclusion was already discussed in the past, e.g., by \citealt{nakar2017}). \\  

{\it Shock breakout}: The first light that the jet generates is released upon breakout of the shock driven by the jet head and the cocoon. The breakout emission depends strongly on the shock velocity \citep{katz2010,nakar2010,nakar2012,levinson2020}, which may vary from ultra-relativistic velocities along the jet axis through mildly relativistic velocities at angles of about $\theta_j$ to sub-relativistic velocities at $\theta \gg \theta_j$. The result is a signal that covers a very large range of frequencies from gamma-rays to optical. Another property that has a strong effect on the shock breakout is whether the shock is parallel or oblique \citep{matzner2013,aartsen2017,irwin2021}. In a breakout driven by a jet, there is a relatively small angular range around the polar angle where the shock is parallel and over most of the stellar edge it is oblique. The observational signature of a jet driven breakout from a stellar surface was not calculated to date.\\

{\it UV/Optical cooling emission:} As the outflow expands, it radiates the part of the internal energy that was deposited by the shocks in what is known as the ``cooling emission". The timescale of this emission of each fluid element depends on its velocity. \cite{nakar2017} provided an analytic estimate of the cocoon cooling emission assuming that the outflow energy is distributed uniformly per logarithmic scale of the proper velocity. This assumption was guided by preliminary numerical results. Here we show that this distribution is universal, and therefore the prediction of \cite{nakar2017} holds true.\\

{\it UV/optical line absorption and emission:} Probably the best observational method to probe the velocity distribution of the ejection is via early spectra. As time passes the expansion of the outflow exposes different layers of the ejecta via line emission and absorption which are imprinted on the UV/optical spectra. Thus, temporally consecutive spectra can provide a "tomography" of the outflow mass and composition as a function of velocity. These methods have been used extensively to study all types of SNe, where the key for probing the fastest moving ejecta is to look at early times, hours to days, before the lines in the fast layers become optically thin. The velocity distribution that we find in this paper predicts that in jet driven explosions  at $v>v_0$ the spectra can be fitted by an ejecta with $\rho(v) \appropto v^{-5} $.\\

{\it Radio emission:} The radio emission is a very sensitive probe of the fastest material since the synchrotron emission from the forward shock driven by this material into the circum-burst medium is bright in the radio. This emission is extremely sensitive to the shock velocity, and thus an explosion with a mildly relativistic outflow is expected to produce a bright radio emission. It is most useful as it enables a reliable measurement of the outflow velocity and a robust lower limit of its energy \citep{pacholczyk1970,chevalier1998,barniolduran2013}. It also provides a rough estimate of the outflow energy which depends on the uncertain microphysical parameters. When the radio light is accompanied by a detectable X-ray emission from inverse Compton radiation, it provides tight constraints on the energy in the shocked region (and a measurement of the microphysical parameters; \citealt{chevalier2006}). In jet driven explosion $E_{MR} \sim E_{SN}$ and therefore we expect to see radio emission that is dominated (at least on a timescale of months to years) by a mildly relativistic material with a comparable amount of energy to that carried by material moving at $v_0$, as measured by the optical SN emission. 

\section{Comparison to observations}\label{sec:observations}

\begin{figure}
	\center
	\includegraphics[width=0.5\textwidth]{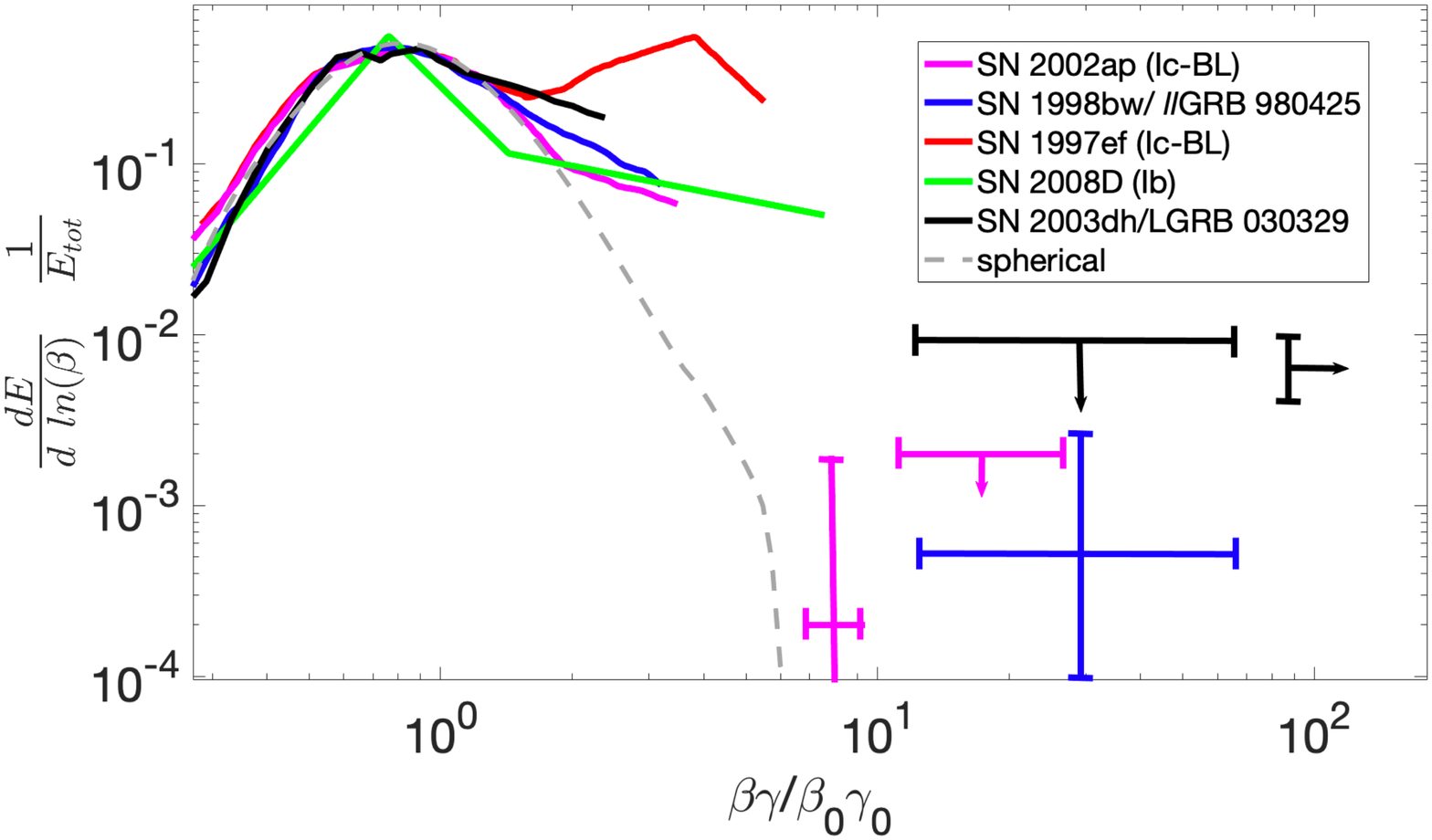}
	\caption{Energy distribution, as derived from observations of a number of stripped envelope SNe and the associated GRB (if there is one). The distributions are all normalized so their peaks coincide at $v_0$. The distribution at low velocities (around $v_0$) were derived based on early SN spectra. The mildly relativistic measurements/limits are based on radio and sometimes X-ray emission (see discussion on the limits of SN2002ap in the text). The ultra-relativistic measurement of GRB 030329 is based on the prompt gamma-ray emission and the afterglow. The distributions are taken from the following references: SN 2002ap: \citet{mazzali2002,berger2002,bjornsson2004}; SN1998bw/ {\it ll}grb980425: \citet{iwamoto1998,li1999}; SN1997ef: \citet{mazzali2000}; SN2008D: \citet{mazzali2008}; SN2003dh/LGRB030329: \citet{mazzali2003,mesler2013}. }     
	\label{fig:observations}
\end{figure}

Fig. \ref{fig:observations} shows the energy distribution as a function of the proper velocity of a number of stripped envelope SNe. One of these is LGRB/SN, one {\it ll}GRB/SN and three are not associated with GRBs\footnote{There were no observed gamma-rays associated with any of these three SNe. In SN 2008D  the possibility of a successful LGRB jet pointing away from us can be rejected at high probability since it is a type Ib SN and we have never seen such SN associated with a GRB. An off-axis jet is also not very likely in SN 2002ap since it took place at a distance of about 10Mpc and the radio emission from such jet must be detected easily as it slows down to mildly relativistic velocities. We have not seen such emission so far, although one cannot rule out the possibility that it will be detected in the future. Finally, in SN 1997ef an off-axis GRB jet cannot be ruled out.}. All distributions are normalized to coincide at the peak at $v_0$, which is around $10,000$ km/s.
The distributions at sub-relativistic velocities (around $10,000$ km/s) are derived by fitting the results of a radiation transfer code to observations of early SN optical spectra \citep{iwamoto1998,mazzali2000,mazzali2002,mazzali2003,mazzali2008}. These distributions provide a good fit to the data, but they are not necessarily unique and therefore their details may deviate from those of the real distributions. Nevertheless, there are two features that are present in all or almost all distributions, which are most likely robust. First, in all SNe, except possibly to SN 1997ef, there is a clear distinctive peak to the energy at $v_0$ ($\approx 10,000$ km/s). Second, in all SNe there is an excess of energy at $\sim 3v_0$ ($\approx 30,000$ km/s) compared to the expectation from a spherical explosion. 

The mildly relativistic ($ v_0 \gtrsim 10v_0 $) limits and measurements were obtained from the radio and X-ray observations \citep{li1999,berger2002,bjornsson2004,mesler2013}. The energy measurements of SN 1998bw/ {\it ll}GRB980425 and SN 2003dh/LGRB03029 are from those references. For SN2002ap the available measurements are of the velocity of the forward shock driven into the circum-stellar medium a week after the explosion, which is $70,000-90,000$ km/s \citep{berger2002,bjornsson2004}. The measurement of the energy at this velocity and of the upper limit at faster velocities depend linearly on the wind density coefficient $A=\dot{M}_w/4\pi v_w$, where $\dot{M}_w$ and $v_w$ are the progenitor's wind mass-loss rate and velocity, respectively. For a typical Wolf-Rayet wind $A \sim 5 \times 10^{11} {\rm g~cm^{-1}}$. In Fig. \ref{fig:observations} we assume that $A$ can be larger or smaller by a factor of 10 compared to this typical value. Regardless of the exact assumptions used to derive the energy from the radio observations, the following result is clear - the mildly relativistic outflow carries orders of magnitude less energy than the sub-relativistic outflow.

The ultra-relativistic energy, which is relevant only to LGRBs, is based on the prompt gamma-ray emission and the afterglow. The best constraints are obtained for LGRBs in which the observations include the entire afterglow, from early times, when the shock is ultra-relativistic, and up to the Newtonian phase that can be observed by late radio emission. Such observations are necessary in order to obtain an energy estimate that is independent of the poorly constrained jet opening angle. In Fig. \ref{fig:observations} we show the estimated jet energy of GRB030329, which is among the GRBs with the best constraints. The ultra-relativistic component in GRB030329 carries $~50$ times less energy than the kinetic energy of the sub-relativistic ejecta SN. \cite{mazzali2014} shows the SN and GRB energies of additional six SNe/GRBs (some LGRBs and some {\it ll}GRBs), in all cases the sub-relativistic SN ejecta carries 10-100 times more kinetic energy than the mildly or ultra-relativistic GRB outflow component.

Another type of comparison between sub-relativistic and mildly to ultra-relativistic kinetic energies appear in \cite{soderberg2006} and \cite{margutti2013}. These studies present a comparison of the kinetic energy around $v_0$ and the kinetic energy of the fastest moving material measured. In some SNe the fastest measured ejecta moves at $\sim 3v_0$, in some it is mildly relativistic, and in some it is ultra-relativistic. In all cases the fastest moving material carries at least one, and typically many, orders of magnitude less energy than the main SNe component at $v_0$. 

A clear conclusion from all these observations is that SNe (associated and unassociated with GRBs) has a prominent peak of the energy distribution at $v_0$ and the energy carried by faster material is lower. Specifically, in most (and possibly all) LGRBs and \llgrbs, the relativistic and mildly relativistic components carry orders of magnitude less energy than the sub-relativistic ejecta. This result leads, based on our results, to a unequivocal conclusion - {\it the energy source of SNe that are associated with GRBs (both long and low-luminosity) is not the GRB jet}. More generally, none of the observed stellar explosions with measured  distribution of the energy at high velocities were generated by jets alone, unless those are choked so deep within the progenitor envelope that the resulting explosion has a clear peak in the energy distribution. This implies that at least in GRB/SNe, there must be at least two types of energy sources at work. One is a narrowly collimated relativistic jet that generates the GRB and another is a much less collimated (possibly, but not necessarily, fully spherical) explosion that drives the SN, where the latter is 10-100 times more energetic than the former. Our simulations show that such a two-component explosion can generate the observed energy distribution. 

Finally, a comparison of the sub-relativistic ejecta energy-velocity distribution, obtained by the early spectra analysis of stripped envelope SNe, with the predictions of a spherical explosion, shows an excess of energy around $3v_0$ that almost certainly indicates a deviation from sphericity \citep{piran2019}. A likely source of the fast moving material in those SNe which are not associated with GRBs, is the cocoon inflated by a jet that was choked not too deep below the stellar surface. Our simulation of a simultaneous jet launching and spherical explosion suggests that the observed distributions can be well explained by such two-component explosions.

\section{Comparison to previous studies}\label{sec:previous}
The first to consider the dynamics of a stellar explosion driven by a relativistic jet were \cite{lazzati2012}. They explored a range of configurations, paying special attention to the effect of the engine duration, which in their simulations affected mostly the depth at which the jets are choked (most of their configurations are of choked jets). They found that GRB-like jets can successfully explode stripped envelope stars and that the amount of relativistic and mildly relativistic material depends on whether the jet is choked or not, and in the former case on the choking depth. All our results are consistent with their findings. 

\cite{barnes2018} carried out a 2D simulation of a jet driven explosion in attempt to generate simultaneously a relativistic component that can produce a GRB and a sub-relativistic ejecta that produces a SN Ic-BL. Based on their simulation they concluded that a jet driven explosion {\it can} produce a GRB/SN that is similar to those that we observe. This is in striking contrast to our conclusion that jet driven explosions are {\it ruled out} as the sources of the observed GRB/SNe. As we explain below the origin of this discrepancy is that \cite{barnes2018} explored only part of the observables predicted by their model. A consideration of the entire range of signals predicted by their model shows that it cannot be the origin of the observed GRB/SNe. 

\cite{barnes2018} performed a simulation of a jet driven explosion that is rather similar to our setup (we discuss some of the differences below). They divided the resulting outflow at the end of the simulation to three components: (i) an ultra-relativistic jet with $\gamma>10$, which in their specific simulation carries about 10\% of the total outflow energy; (ii) a sub-relativistic ejecta with $\beta<0.2$, which carries 40\% of the energy; and (iii) the rest of the ejecta ($0.2<\gamma\beta<10$) that contain 50\% of the outflow energy. These results are consistent with our simulations. The energy fraction that goes into the ultra-relativistic jet is set by the exact choice of their engine luminosity and duration, and the rest of the energy is divided roughly equally between the sub-relativistic and mildly relativistic outflow. 

In the next step, they assumed that the relativistic jet can generate a GRB, and in order to test if the outflow can also generate a SN, they carried out a radiative transfer simulation where they considered only the sub-relativistic ejecta. Namely, in their paper they ignored the observational signature of the material with $0.2<\gamma\beta<10$ altogether. As discussed above, a mildly relativistic outflow with energy that is comparable to that of the SN has a clear observational signature (e.g., in the radio), which is not seen in any of the observed GRB/SNe. Since in all observed cases the mildly relativistic material carries orders of magnitude less energy than the SN ejecta, the model presented by \cite{barnes2018} is inconsistent with observations of GRB/SNe. In a follow-up study, \cite{shankar2021} carried out several simulations similar to that of \cite{barnes2018}, varying the properties of the jet (opening angle, work time, etc.). Similarly to \cite{barnes2018}, they focused only on the emission from material with velocity $<0.2$c and did not consider the observational signatures of mildly relativistic material. 

There is another point raised by \cite{barnes2018} that merits a discussion in the context of our results. The jet in \cite{barnes2018} is launched with a luminosity that decays exponentially, $L_j = L_0 \exp[-t/\tau]$ where $L_0$ is a constant and $\tau=1.1$s is a characteristic decay timescale. The jet is successful, its breakout time is $t_b \approx 6$ s and the length of the relativistic outflow that breaks out successfully, $\Delta R$, satisfies $\Delta R/c \approx 6.7 s \gg \tau$. In their paper, \cite{barnes2018} argued that this implies that the GRB duration is detached from the duration of the engine activity and that $\Delta R/c$ is affected by a stream that is generated via conversion of cocoon pressure to kinetic energy. We think that the explanation is different and that in fact $\Delta R/c$ is a direct reflection of the engine activity. The engine continues to work (although at a lower luminosity) also at $t>\tau$, and the relativistic outflow that breaks out of the star is launched at $t>\tau$. More precisely, it is launched at $t>t_b-R_\star/c \approx 3$s. This is consistent with the fact that the energy in the relativistic outflow is $\int_{3s}^\infty L_0\exp[-t/\tau] dt$, which is 10\% of the total injected energy, $L_0 \tau$. Thus, $\Delta R$ is determined by the launching of the jet at $t>3$ s and the time where the jet luminosity drops to the point that it cannot penetrate through the highly-pressurized cocoon anymore, which happens roughly at $10$s. The result is $\Delta R \approx 7$s, and if this outflow generates gamma-rays, then the signal duration does correspond to the time during which the engine launches the part of the jet that breaks out of the star, and it reflects the jet engine luminosity during that time.

\section{Conclusions}\label{sec:conclusions}
We explored analytically and numerically the velocity distribution of jet-driven explosions and the observational signatures that this distribution entails. We also performed a single simulation of an explosion with two energy sources, which are active simultaneously -- a relativistic jet and a spherical explosion. Our main conclusions are:
\begin{itemize}
    \item An explosion driven by an unmagnetized or weakly magnetized successful jet produces an outflow distribution with a constant amount of kinetic energy per each logarithmic scale of proper velocity ($\gamma\beta$). In a successful jet, the range of the flat distribution is between $v_0=\sqrt{E_c/M_\star}$ and $\gamma \approx 3$. This result is universal with no dependence on the jet or progenitor parameters that we explored, which cover most of the parameter phase-space of GRBs.
    
    \item The distribution of the outflow is not isotropic. Ejecta with $\gamma \sim 2-3$ is confined to an opening angle of $\sim 30-45$ deg around the jet axis. The opening angle increases for lower velocity ejecta and for $v \lesssim 0.1$c the outflow is quasi isotropic.
    
    \item The jet itself composes a separate narrowly collimated relativistic component with\footnote{In this paper we did not discuss the angular structure of the jet Lorentz factor that varies from a maximal value along the axis to $\gamma \sim 3$ at the interface between the jet and the cocoon. See \cite{gottlieb2020,gottlieb2021} for a detailed discussion of the jet structure.} $\gamma>3$. The jet energy, $E_j$, can be, in principle, much larger or much smaller than that of the expanding envelope, $E_c$, as it depends on the duration of the engine after the jet breakout. However, for typical LGRBs we expect $E_j \sim E_c$.
    
    \item In explosions driven by collimated choked jets, the outflow energy distribution is also flat in log space of $\gamma\beta$. Similarly to successful jets, the flat distribution starts at $v_0$, but unlike successful jets it has an upper cutoff at $\gamma\beta < 3$. The cutoff velocity  decreases with the depth at which the jet is choked. 
    
    \item The velocity distribution of an outflow from an explosion driven by a highly magnetized jet is expected to be rather similar to the one of unmagnetized and weakly magnetized jets. It may be exactly the same if the jet magnetic field is dissipated at the nozzle that forms near the jet base. Otherwise, the outflow from the outer cocoon, which ranges over a decade of velocities in the sub-relativistic regime, is similar to that of an explosion driven by an unmagnetized jet, whereas the outflow from the inner cocoon may generate a distribution at mildly and possibly ultra-relativistic velocities that is not flat. However, even in that case the mildly (and possibly ultra) relativistic outflow from the inner cocoon carries a comparable energy to the sub-relativistic component.
    
    \item The velocity distribution of a jet driven explosion can be identified via a number of different observational channels. In the UV/optical the fast moving material is predicted to generate an early flash (on timescales of minutes to days) powered by its cooling emission \citep{nakar2017}. It can also be seen in this frequency range via absorption during the first several days \citep{piran2019,izzo2019}. Finally, the mildly relativistic (and possibly ultra-relativistic in case of highly magnetized jet) component of the outflow generates a bright radio, and possibly also X-ray, emission that can be seen for a long time (possibly even years) after the explosion. 
    
    \item A two-component explosion that includes an energetic quasi-spherical energy deposition and a less energetic jet produces an outflow with a velocity distribution that has a prominent peak that corresponds to the spherical component and a fast tail that corresponds to the jet. The energy ratio between the peak and the tail corresponds to the energy ratio between the spherical component and the jet. This result shows that a choked jet can generate the excess in high velocity material seen in stripped-envelope SNe, if it accompanies a more energetic quasi-spherical explosion, as suggested first by \cite{piran2019}.
    
\end{itemize}

An examination of the available data on the velocity distribution in GRB/SNe shows that in all cases the mildly and ultra-relativistic outflow carries at least an order of magnitude, and in most cases several orders of magnitude, less energy than the sub-relativistic SN ejecta. This is true both for LGRBs and for {\it ll}GRBs. Therefore, our results imply that {\it GRB/SNe of all types are not generated by a jet-driven explosion. Instead the explosion mechanism of LGRBs and \llgrbs must include two components. A wide angle (possibly quasi-spherical) energy deposition that ejects the sub-relativistic SNe outflow and a narrowly collimated relativistic GRB jet that carries 1-10\% of the total outflow energy in LGRBs and possibly (but not necessarily) less in llGRBs.}

Interestingly, the same energy ratio between two similar components was observed in the outflow from the BNS merger GW170817, although each one had $\sim 10$ times less energy than in LGRBs. The sub-relativistic quasi-spherical ejecta carried $\sim 10^{51}$ erg, whereas the narrowly collimated ultra-relativistic jet carried $10^{49}-10^{50}$ erg \citep[e.g.,][and references therein]{nakar2020,margutti2021}. In GW170817 the common picture is that the sub-relativistic outflow is driven mostly by winds from a $\sim 0.1 {\rm~M_\odot}$ accretion disk that forms around a newly formed central BH, and the jet is generated during the accretion process (see review by \citealt{nakar2020}). In collapsars it is reasonable to expect that the total accreted mass, and thus the available energy, is ten times larger, i.e., $\sim 1 {\rm~M_\odot}$. Therefore, this analogy supports a picture where the quasi-spherical energy source in collapsars is the wind from the accretion disc, and the jet is launched as part of the accretion process on the central compact object.

\section*{Acknowledgements}
This research was partially supported by a consolidator ERC grant 818899 (JetNS) and by an ISF grant (1114/17). OG is supported by a CIERA Postdoctoral Fellowship. Computational support was provided by the NegevHPC project.

\section*{Data Availability}
	
	The data underlying this article will be shared on reasonable request to the corresponding author.

\bibliographystyle{mnras}
\bibliography{SNeJet} 

\appendix

\section{Comparison to 3D simulation}\label{sec:3D}
\begin{figure}
	\center
	\includegraphics[width=0.5\textwidth]{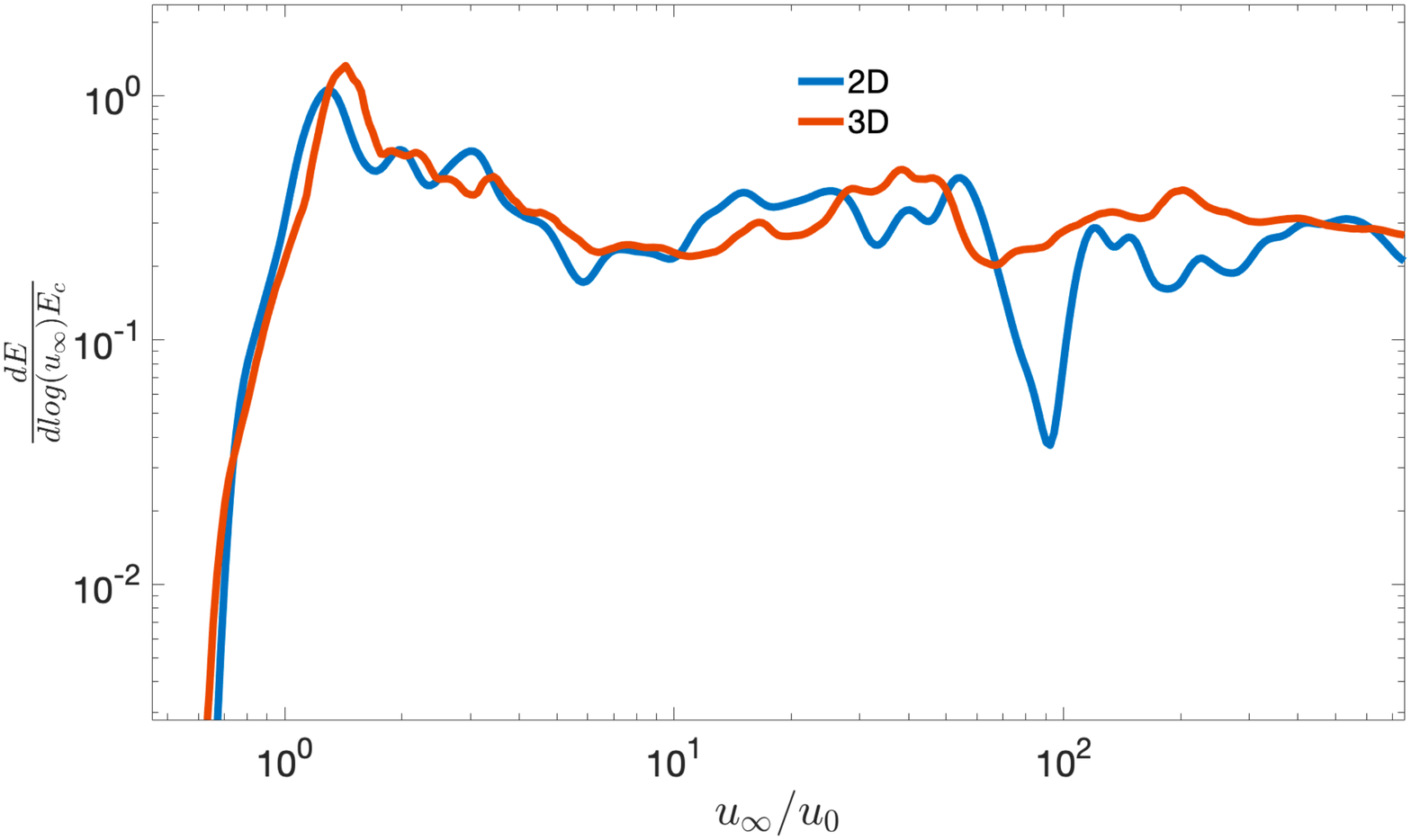}
	\caption{A comparison of 2D and 3D simulations of the same setup. The normalized energy distribution as a function of $u_\infty=\sqrt{h^2\gamma^2-1}$. The simulations run until the jet head reaches 10$R_\star$.    
	}
	\label{fig:2D_3D}
\end{figure}
To verify that our 2D simulations are valid for the specific purpose of this paper, we compare one 2D run to a similar 3D simulation. Since 3D simulations at resolution that is high enough to resolve the details of the head and the jet structures are computationally expensive, we could not run the 3D simulation until the entire star explodes, and all the envelope reaches the homologous expansion. Instead, we use 3D and 2D simulations of model $L_c$ from \cite{gottlieb2021} that run until the jet head reaches 10$R_\star$ (all the setup and simulation details can be found in that paper). At this time, not all the stellar envelope is shocked and the velocity distribution is not expected to be identical to the one seen in Fig. \ref{fig:allJets}. However, at later times the jet no longer deposits energy into the cocoon, and the 2D and 3D evolution are expected to be similar. Thus, if the 2D and 3D simulations provide similar results when the jet head reaches 10$R_\star$, they are expected to be consistent when the entire star explodes as well. Fig. \ref{fig:2D_3D} depicts a comparison of the energy distribution in 2D and 3D. Since not all the material had enough time to expand at that time, we plot the energy distribution as a function of $u_\infty=\sqrt{h^2\gamma^2-1}$, as an approximation to the final proper velocity of each fluid element during the homologous phase. Fig. \ref{fig:2D_3D} shows that the 2D and 3D simulations are in an excellent agreement, implying that 2D simulations are valid.

\section{Convergence test}\label{app:convergence}
\begin{figure}
	\center
	\includegraphics[width=0.5\textwidth]{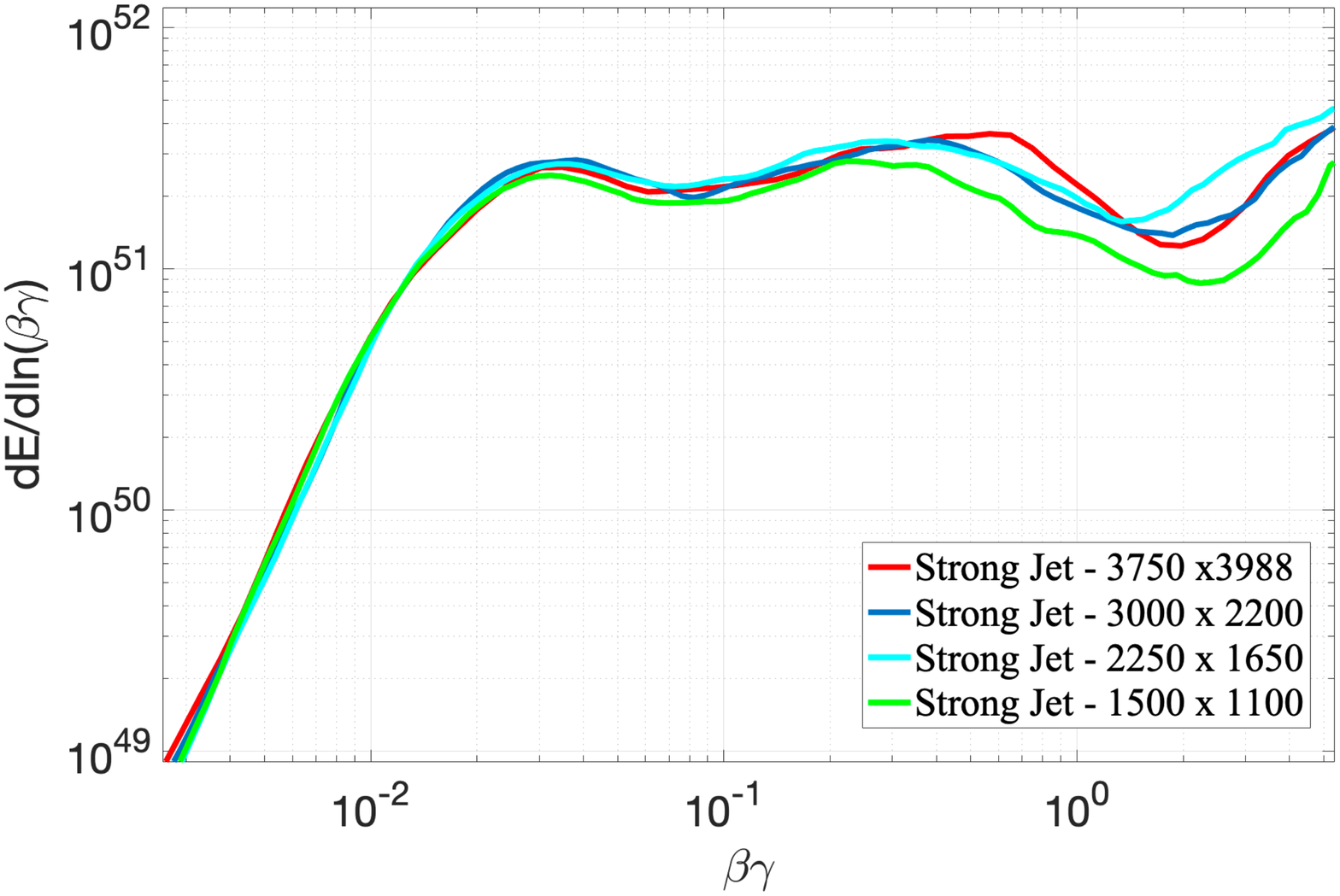}
	\caption{Convergence test. The energy distribution as a function of the proper velocity during the homologous phase of four simulations of the strong jet configuration in different resolutions.    
	}
	\label{fig:convergence}
\end{figure}

Numerical convergence is verified by carrying out four simulations of the "strong jet" setup with different resolutions. In addition to our canonical resolution, $ 3000 \times 2200 $, we preform two lower resolution simulations, in which we increase the dimensions of each cell by a factor of 2 [4/3], to a resolution of 1500x1100 [2250x1650]. We also carry out a simulation with a higher resolution and a different aspect ratio to verify that our results are not affected by it. We use 3750 grid points in the $ \hat{z} $-direction: 1250 points are uniformly distributed from $z=Z_i=1 \times 10^{9} \,\mathrm{cm}$ to $z=1 \times 10^{11} \,\mathrm{cm}$, and the rest 2500 points have a logarithmic distribution from $z=1 \times 10^{11} \,\mathrm{cm}$ to $z=1 \times 10^{13} \,\mathrm{cm}$; and 3988 grid points in the $ \hat{r} $-direction: 250 points are uniformly distributed from $r=0$ to $r=1 \times 10^{9} \,\mathrm{cm}$, another 1238 points have a logarithmic distribution from $r=1 \times 10^{9} \,\mathrm{cm}$ to $r=1 \times 10^{11} \,\mathrm{cm}$, and the rest 2500 points have a logarithmic distribution from $r=1 \times 10^{11} \,\mathrm{cm}$ to $r=1 \times 10^{13} \,\mathrm{cm}$. 

Fig. \ref{fig:convergence} shows the energy distribution as a function of the proper velocity during the homologous phase of the four simulations. All simulations show a roughly flat distribution in the range $\gamma\beta \approx \beta_0-3$, with rather minor differences. At lower velocities, $\lesssim 0.1c$, all simulations are converged. At higher velocities, the lowest resolution simulation slightly deviates from the rest of the simulations, which are very similar to each other at all the velocities. We conclude that our canonical resolution is high enough to study the velocity distribution of the outflow at $\gamma \lesssim 3$.
\end{document}